# Colloidal Quantum Nanostructures: Emerging Materials for Display Applications

*Yossef E. Panfil,[a] Dr. Meirav Oded,[a] Prof. Uri Banin*[a]*

*Dedicated to the employees of Merck and Qlight working in the field of quantum materials for display applications*

[a]    Institute of Chemistry and the Center for Nanoscience and Nanotechnology

The Hebrew University of Jerusalem, Jerusalem, 9190401 (Israel)

E-mail: uri.banin@mail.huji.ac.il

**Colloidal quantum nanostructures** constitute outstanding model systems for investigating size and dimensionality effects. Their nanoscale dimensions lead to quantum confinement effects that enable tuning of their optical and electronic properties. This Review presents current and potential applications of semiconductor nanocrystals as sophisticated materials for display technologies.

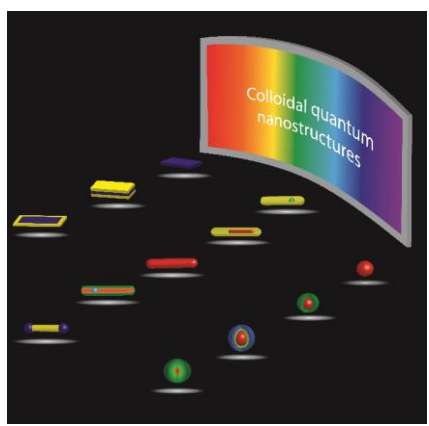

Abstract: Colloidal semiconductor nanocrystals (SCNCs) or, more broadly, colloidal quantum nanostructures constitute outstanding model systems for investigating size and dimensionality effects. Their nanoscale dimensions lead to quantum confinement effects that enable tuning of their optical and electronic properties. Thus, emission color control with narrow photoluminescence spectra, wide absorbance spectra, and outstanding photostability, combined with their chemical processability through control of their surface chemistry leads to the emergence of SCNCs as outstanding materials for present and next-generation displays. In this Review, we present the fundamental chemical and physical properties of SCNCs, followed by a description of the advantages of different colloidal quantum nanostructures for display applications. The open challenges with respect to their optical activity are addressed. Both photoluminescent and electroluminescent display scenarios utilizing SCNCs are described.

1. **Introduction**

Display devices that enable the input and output of information are central to the digital information era. We use displays everywhere and a lot of the time - at home, in the workplace, and on the go. Each of us has at least one in our pockets. Over the past 100 years, revolutions and evolution of display device technologies have been closely dependent upon the generation of innovative materials for displays enabled by clever chemistry. The focus of this Review is colloidal semiconductor nanocrystals (SCNCs), also termed colloidal quantum nanostructures herein, which constitute an emerging family of materials for display applications.

SCNCs are made up of tens to tens of thousands of atoms arranged in the ordered structure of the bulk semiconductor from which they are derived.[1-3] Their nanoscale dimensions place them as an intermediate form of matter between molecules and bulk macroscopic crystals. Controlling the size, shape, and composition, as illustrated in the series of SCNCs in Figure 1, affords powerful control over their optoelectronic properties.[4-7] Moreover, they are synthesized by wet chemical approaches that allow their manufacture on a large scale. The surface of the nanocrystals is inherently overcoated by a layer of surface ligands, thus enabling highly flexible manipulation for their incorporation from solution,[8] for example, to print individual pixels for displays. The flexibility in surface chemistry manipulations also allows for their incorporation into different matrices, for example, in polymer films that can be seamlessly integrated into a display.

Colloidal SC quantum dots (CQDs), the first type of SCNCs, were conceived in the early 1980s and have been developed and perfected over the past 25 years, both in terms of synthetic control over the materials characteristics and the understanding of their optoelectronic properties. A well-known property of CQDs is that the emission color of the nanocrystals can be tuned merely by changing their size, which has become a hallmark of nanoscience (Figure 2a). This is a result of the quantum confinement effect,[9,10] and it provides vivid colors that are a perfect match for the demanding needs of current and next-generation displays.

The images in a display are made up of a large number of small pixels with individually controlled light output. Adjacent pixels of typically three base colors, red, green, and blue (RGB), are used to generate color images. Figure 2b depicts the International Commission on Illumination (CIE) chromaticity diagram. This diagram, composed of the base RGB colors, maps the colors which are visible to the human eye ($\lambda$=380-780 nm) in terms of the hue and saturation. Each set of three colors located on the map defines a triangle. The colors enclosed within that triangle can be generated by a suitable combination of different ratios of its corners. Increasing the purity of the colors in the corners thus results in larger coverage of the CIE chart by the triangles, which is termed the color gamut. Sunlight spreads the entire color gamut, while the RGB spectra in most commercially available displays today spread only the black triangle (s-RGB standard).

SCNCs are characterized by a narrow emission spectrum (Figure 2) with high fluorescence quantum yield (QY), thus providing an obvious advantage for their implementation in display applications to yield a wider representation of the color gamut. The CQDs are also characterized by a wide absorption spectrum spanning an energy range larger than their band gap energy, and have high photostability, thereby opening an opportunity for their implementation in displays using optical excitation. An additional option is to excite the SCNCs by charge injection, and CdSe NCs were already incorporated in 1994 into semiconducting polymers in hybrid organic/inorganic light emitting diodes (LEDs).[11] Since then, SCNCs have been used in down-converting backlight applications in LCDs as well as in electroluminescent displays.[12-16]

Besides using size to control their optical and electronic properties, SCNCs can be synthesized in a variety of shapes and dimensionalities (see below);[17-22] these characteristics govern the polarization and the directionality of the emission of the NCs. The surface of the NCs is easily processed, and hence NCs can be introduced onto various phases/substrates such as hydrophobic or hydrophilic media, and onto or into flexible or rigid substrates.[23-26] These qualities also make NCs appealing in a variety of additional applications, such as sensing,[27] imaging and bio-labeling,[28-30] solar cells,[31, 32] lasing,[33, 34] photocatalysis,[35] and printed electronics.[15,36]

In this Review, we will discuss the utilization and potential of SCNCs in the field of displays. We briefly describe key principles for the synthesis of SCNCs, and then provide the fundamental physicochemical background regarding the energy-band structure and the charge-carrier dynamics of the NCs. We then describe the two basic set-ups available for display applications: LCD-based photoluminescence and LED-based electroluminescence (EL). We conclude with perspectives on what the future holds for SCNCs in display applications.

## 2. Synthesis of SCNCs

SCNC synthesis has evolved into an important field of materials, inorganic, and solid-state chemistry. Numerous reviews address this topic in great detail and herein we only briefly describe the key principles.[42, 43] The "ideal" SCNC sample for display applications should be characterized by a well-defined and controlled size, with a narrow size distribution to achieve the stringent color emission requirements of displays. A high degree of crystallinity is also important, since defects can serve as trap sites that quench the fluorescence. Good surface passivation is also required to eliminate surface traps.

Although early synthetic approaches for II-VI SCs employed water-based synthesis,[2] the most successful and robust methods use high-temperature synthesis in organic media.[44,45] The separation of nucleation and growth is one essential principle for achieving the monodisperse samples. In this way, all the nanocrystals are "born" together in a well-defined nucleation step and then grow together to yield a narrow size distribution upon quenching the growth process. A second central concept is that of focusing the size distribution,[46-49] and in general avoiding Ostwald ripening, by which the large particles in the distribution grow on account of the dissolution of the smaller less-stable particles leading to a broadening of the size distribution. Realizations of these principles are addressed by the "hot injection" method, in which suitable reactive precursors are rapidly administered to a vigorously stirred high-temperature boiling solvent, which often itself also coordinates (or, in the presence of coordinating ligands) and arrests the growth to the nanoscale regime. An excess of one or both precursors of the SCNC is added to suppress the Ostwald ripening effect and further additions of the precursors are administered to grow the seeds to the desired sizes while maintaining the distribution in the regime of size-focusing.

Other approaches have also been developed that may be even better suited to the batch synthesis of industrial amounts of NCs. One such approach is the "heating-up" method, in which a "one-pot mixture" of the solvent and the reactive precursors is stirred at low temperatures and then heated-up to induce the crystallization reaction.[50-52] It was also reported for metal-sulfide NPs, that working at high concentrations, near the solubility limit of the precursors, results in reduced sensitivity to variations in the reaction parameters, which results in the reproducible and large-scale production of monodisperse NCs.[53] Unlike the hot-injection technique, in which the size of the particles can be increased during the synthesis, while maintaining the size dispersity of the NCs, each reaction has a fixed set of conditions in this method (i.e. solvents with different boiling points, different ratios between the precursors, and time), which yields specific size-focused NCs without the need for additional size-selective purification steps.

These methods can be applied to a variety of semiconductor materials to form well-controlled NCs. The best-developed systems are those of cadmium chalcogenides, in particular, CdSe, which span the visible range for displays. The synthesis of shape-controlled SCNCs is also highly advanced in the II-VI semiconductor materials.[17,54,55] Lagging behind are the III-V semiconductor materials, especially InP, which can be synthesized by related approaches.[18,56-58] In addition, the method enables the growth of shells of other SCs on the NCs by similar principles to create heterostructures, which is essential for the further improvement and control of the optoelectronic characteristics of the SCNCs.

## 3. Quantum Confinement, Heterostructures, and Shape Effects in SCNCs

### 3.1. Optical and Electronic Properties

The energy levels of SC bulk materials are composed of two continuous bands separated by an energy gap ($E_g$): a fully occupied lower valence band (VB) and a higher empty conduction band (CB). Excitation of an electron from the VB to the CB leaves behind a hole with a positive charge in the VB. The electron-hole pair thus formed interacts through Coulomb attraction to form an exciton with a characteristic binding energy and an exciton Bohr radius of $a_0$, derived in analogy to the Bohr model of the hydrogen atom. As a consequence of the typically light effective masses of electrons and holes within the SC, and because of the dielectric screening, the binding energies of excitons in typical SCs of relevance for display applications are on the order of or smaller than the thermal energy, which leads typically to dissociation of the exciton in the bulk phase. When one or more dimensions of the SC nanostructure is reduced and become comparable or lower than the Bohr radius, strong quantum confinement effects govern the electronic level structure and lead to the discretization of the levels and to an increase in the band gap as the size decreases.[59]

A "particle in a spherical box" model can be used to describe the discrete energy levels of CQDs (Figure 3a). Since the potential in the spherical geometry depends solely on the radius, just like in the hydrogen atom, the spherical harmonics describe the angular wave functions of the confined states, where the ground states in either the CB or VB will be $1S_e$ for the CB and $1S_h$ for the VB. The next levels will be $1P_{e/h}$, $1D_{e/h}$, and then $2S_{e/h}$. Thus, in CQDs, the band gap transition becomes that of the $1S_e1S_h$ exciton state and the discrete level structure leads to emission of a photon with a narrow PL peak blue-shifted upon reducing the CQD radius.

Figure 3b depicts the dependence of the optical characteristics of the QD on the size and dimensionality.[60] The energy gap increases for strongly quantum confined systems. This is manifested in blue-shifted absorption and emission spectra.[61] As the QD size is reduced in all dimensions, the quantum confinement increases. 2D or 1D systems confined with a similar critical size (e.g. thickness in the 2D system or diameter in 1D) will have lower $E_g$ values than 3D confined systems. The effects of dimensionality will be discussed more thoroughly later on.

Direct characterization of the energy levels within the bands was convincingly demonstrated by scanning tunneling microscopy (STM) measurements of single CQDs on a conducting substrate. Following imaging to locate the CQD, the tip is held at a fixed height above the particle and the scanning tunneling spectrum is obtained by measuring the current-voltage (*I-V*) characteristics, as shown in Figure 3c for an individual InAs CQD.[62, 63] The density of states was extracted by plotting the tunneling conductance spectrum (d*I*/d*V* versus *V*, termed scanning tunneling spectroscopy, STS, Figure 3d). The CQD energy gap is clearly resolved around zero bias. On the positive bias side, which corresponds to CB states, a doublet appears after current onset, followed by a group of six adjacent peaks. The first doublet is attributed to tunneling through the lowest state in the conduction band, $1S_e$, which is indeed doubly degenerate. The energy spacing between the two peaks is due to the charging energy ($E_C$) for each tunneled electron. Similarly, the next group of six peaks corresponds to the charging of the second CB level, $1P_e$. The atomlike character of the levels, along with a consecutive electron filling ("Aufbau principle"), demonstrate the artificial atom behavior of the QD. A similar characterization was performed for the negative bias side, which relates to tunneling through the filled VB levels. The characteristic peak structure of the VB is denser due to the more elaborate VB band structure of the bulk semiconductor material, and to mixing of the levels.

Examination of the STS spectra of a series of different size InAs CQDs (Figure 3e) clearly demonstrates the quantum confinement effects. As the size decreases, a systematic increase in the energy gap is seen. On the CB side, current is typically onset by a doublet of peaks corresponding to the $1S_e$ state and followed then by the higher order multiplet of the $1P_e$ state. Size reduction also increases the spacing between the $1S_e$ and $1P_e$ states, and the charging energy is also increased.

### 3.2. Charge Carrier Dynamics

The most common pathways for the generation of excitons in SCNCs, which are highly relevant for display applications, are either by optical excitation, which is used in photoluminescence applications, or excitation by electrical charge injection, which is used in EL applications. In optical excitation, the absorption of an energetic photon by the NC leads to the creation of an exciton (Figure 4a). In electrical excitations, charge injection to CQDs occurs through electron- and hole-transporting layers (ETL and HTL, respectively). When an electric bias is applied, current flow takes place and electrons are injected into the CB of the QD through the ETL, while electrons are also drawn from the VB through the HTL, thereby leading to the transportation of holes into the VB and to the creation of excitons in the QD (Figure 4a).

When a "hot exciton" (an exciton with energy way above the band gap) is created, both the electron and the hole first relax to the band gap states in a thermalization process involving electron-phonon and electron-hole interactions (Figure 4b). The relaxation usually occurs on a time scale ranging from sub-pico- to a few picoseconds.[64-67] Once the electron and hole reach the ground states of the CB and VB, respectively, they can recombine either in a radiative manner or in a nonradiative manner. The nonradiative recombination of the exciton is clearly undesired for display applications and occurs either by trap-assisted recombination[8] or by Auger recombination.[68]

An example of trap-assisted recombination is illustrated in Figure 4c: an electron decays to a trap state which is located within the CQD band gap, from which the electron can decay to the VB and recombine with a hole. The process can be approximated by a two-step decay process, with an overall time scale of nano- to microseconds.

Auger-type recombination (Figure 4c) is a three-particle process, in which the energy released by the recombination of one exciton is used to excite another charge carrier (electron or hole) above the band gap, with a time scale of about 100 picoseconds. Auger-type recombination can occur with charged CQDs or when multiexcitons are generated (i.e. upon high flux irradiation, injection of multiple charge carriers, or upon either the electron or the hole entering a trap state). Therefore, it is more common in electrical excitations where charging effects are more likely to occur.

For display applications, the radiative channel is exclusively preferred, since high QYs are essential. Moreover, when the excess energy of an exciton is transferred to a nonradiative recombination channel, apart from losing a potentially emitted photon, the NC can enter an "off" state, in which the NC is optically inactive. This phenomenon, termed "blinking", will be discussed in more detail in Section 3.5.

To exclude all nonradiative processes, which quench the fluorescence of the NC, we need to understand the origin of each of the nonradiative routes. Atoms not fully coordinated in the SCNC (either on the outer surface or inside the NC) can act as traps for the trap-assisted nonradiative recombination process. Different types of NCs promote different trap states. For example, core-only SCNCs, which are passivated solely by organic molecules, quite often demonstrate low QY values, because of traps on the surface. The organic molecules are considered "poor" passivators, as they cannot fully passivate all the dangling bonds at the surface as a result of steric hindrance and chemical binding considerations, since the ligand will have affinity to either the positive or negative sites. Therefore, in an effort to improve the passivation and maximize the PL QY, core-shell heterostructures of SCNCs were developed, as discussed in the following section.[69]

### 3.3. Heterostructured SCNCs

Heterostructure SCNCs composed of two (or more) separate areas of SC materials, such as core-shell structures, are essential for tailoring SCNCs for display applications. In core-shell particles the SC core is surrounded by a passivating shell of a second SC, thus allowing the potential energy profile for the charge carriers to be shaped. The optical and electronic properties of the SCNC can be controlled by the band alignments of the core and shell SCs (Figure 5).[6,70-73] Systems in which the band gap of the core and the shell are straddled, such as is the case for CdSe/ZnS, are denoted as type I systems (Figure 5). In this case, the core is electronically passivated and both the electron and the hole are localized within the core. In these systems, the inorganic shell passivates the traps on the core/shell interface, which typically leads to a high QY and enhanced stability.

Systems in which the band gap of the core and the band gap of the shell are staggered, such as CdSe/ZnSe, are denoted as type II systems (Figure 5). In this case, the first exciton transition is red-shifted relative to the band gap of either component and both charge carriers are localized in separate regions of the NC (either core or shell).[74] This offers access to wavelengths that are not possible in either SC structure separately. Although type II systems are usually more suitable for applications which require charge separation (such as photocatalysis or charge-transfer processes), elevated external quantum efficiencies for a type II system have been reported.[12]

An interesting intermediate case is the quasi type II system. The band alignment of the core and the shell are straddled, but the band offset of one of the bands (CB or VB) is small. An example of such a system is the well-studied CdSe/CdS system, in which the hole is confined to the CdSe core, but the electron wave function can also delocalize into the shell. In such a case, the first exciton transition is also red-shifted upon growth of the shell.[75]

SCNC core/shell systems of ZnS- capped CdSe NCs were presented in 1996 which showed a QY of 50% at room temperature, an increase of an order of magnitude in the QY compared to the bare CdSe core.[76] Shortly after, a report was published on highly fluorescent core-shell CdSe/CdS CQDs.[75] Since then, numerous publications have appeared on a variety of different core-shell structures, including different types (type II, reverse type I, quasi type I, and type II),[6,77,78] multiple shell growth,[79] and in different synthesis approaches (by successive ion-layer adsorption and reaction[80] and from a single precursor[81]).

However, the core and shell materials have different lattice parameters; hence, structural defects can occur at the interface which serve as traps and provide a nonradiative decay channel, which again reduces the QY of the NC. Therefore, apart from considerations of band alignment, the appropriate shell must be tailored for a minimal lattice mismatch between the core and the shell to avoid structural defects. An interesting solution for the lattice-mismatch problem is the use of an alloy or graded shell as a mediating layer. In this structure, an interfacial alloy layer, which gradually changes from one material to another, is used to ease the strain arising from lattice mismatch. The concept was first introduced in 2005 on a system of highly luminescent CdSe/CdS/Zn$_{0.5}$Cd$_{0.5}$S/ZnS multishell NCs.[37] High QY values were achieved by gradually changing the shell composition in the radial direction from CdS, which has a quasi-type II band alignment but lower lattice mismatch with CdSe, to ZnS, which has high band offsets but large lattice mismatch with CdSe. This concept was widely implemented on SCNCs of various types and materials. Apart from enabling the engineering of different shells with dissimilar lattice parameters, the graded alloy layer also decreases the decay rate through nonradiative Auger recombination. This effect and its influence on the PL and QY of NCs will be discussed in detail in Section 3.5.

### 3.4. Dimension and Dimensionality Effects

Besides the development of the zero-dimensional CQDs and spherical core/shell nanocrystals, elongated quasi-1D CdSe rods were discovered, which paved the way towards shape control of SCNCs, offering further selection of important properties for display applications.[17] In 2003 a dot-in-rod architecture of a CdSe dot within a CdS rod was introduced,[82] and later a seeded growth synthesis for this interesting system was reported.[83,84] This CdSe/CdS dot-in-rod heterostructure system has a mixed dimensionality as it integrates a 0D confined seed within a 1D confined system. Since then, SCNCs with versatile compositions, dimensions, and sometimes also different dimensionalities within one particle have been synthesized.

A first property important for display applications is illustrated in the CdSe/CdS seeded rod, where the rod component has a significantly larger volume than the seed, and as a result, the overall absorption is dominated by this component. Figure 6a shows the absorption and emission spectra for a CdSe/CdS seeded rod system (core diameter 4.3 nm, rod diameter 4.7 nm, and rod length 30 nm), and the corresponding absorption and emission spectra of the bare CdSe cores. For the mixed system, absorption features with wavelengths higher than $\lambda=500$ nm are associated with electronic transitions involving the CdSe core states, whereas for wavelengths shorter than $\lambda=500$ nm, a sharp rise in the absorption is observed, which is associated with electronic transitions involving the CdS rod states. The emission in seeded rods emanates from the band edge, which corresponds to the lowest excitonic state of the CdSe seed. The direct consequence of the difference in the dimensions of the parts constituting the particle is that the large-volume component of the rod acts as an efficient antenna for absorption because of its significant extinction coefficient, while the emission is related to the seed, which can be engineered to emit in a desirable wavelength. This characteristic is important for display applications and specifically for LCD displays using a film containing SCNCs which down-convert blue light into red and green colors (see Section 5). The large absorption cross-section of the SCNCs in the blue region of the spectrum, which is attributed to the large rod component of the particle, efficiently absorbs the blue light of the LED and converts the light into the desired specific green or red color point.

Another manifestation of the importance of the separation between the absorption and emission of SCNCs that is particularly demonstrated by the CdS/CdS dot-in-rod structure was reported for ink-jet printing fluorescent layers containing SCNCs.[85] The spherical CdSe/CdS dots suffer from self-absorption effects, in which particles absorb the light emitted by other particles because of significant overlap between the absorption and the emission spectra of these NCs. This self-absorption causes the effective external emission QY to decrease significantly and induces changes in the fluorescence color by shifting the emission energy to longer wavelengths, whereas, for dot-in-rod structures, the QY remains at high levels and the emission peak does not red-shift, even in a highly concentrated layer.

In the close-packed SCNC layers required for thin light-converting structures, or in LEDs, an additional possible loss mechanism is related to Förster-type fluorescence resonance energy transfer (FRET), in which an excited NC serves as an exciting donor through nonradiative dipole-dipole interactions to a neighboring NC that serves as an acceptor. The FRET process will also lead to an undesired red-shift of the PL and a reduced QY. Dot-in-rod structures exhibit significant improvement over spherical ones in such close-packed fluorescent layers, since their geometry leads to a reduced proximity between the emission centers and hence to near-elimination of the FRET processes (Figure 6b). Such SCNC systems show promise as RGB subpixels in display applications with inkjet printing or by additional patterning methods.

SCNC heterostructures comprised of different shapes and dimensionalities also offer advantages and opportunities in the facilitation of charge injection or extraction of both carrier types from different regions. One kind of such structure is a dumbbell-shaped NC. Unlike core-shell NCs, where at least one of the charge carriers is in an inaccessible region (irrelevant if it is a type I or type II NC), in a dumbbell-shaped NC, both of the charge carriers can be found in directly accessible regions. DHNR (double heterojunction nanorod) dumbbells comprised of CdSe tips at the end of a CdS rod and over-coated with a ZnSe shell,[40] were found to be very efficient both as an emissive layer for electroluminescent QD LED displays and as a light detector.[12] This opened the possibility of multifunctional displays (see Section 7). Recently, ZnTe/ZnSe nanodumbbells (NDBs) were presented where the fluorescence could be tuned between about λ=500-585 nm by changing the ZnSe tip size,[87] which can also be a promising Cd-free system for use in multifunctional displays.

The SCNC shape also has major implications on the polarization properties of the SCNCs emission, which is also applicable to display applications. Since the early studies on CdSe nanorods[88, 89] and InP nanowires,[90] it has been well-known that elongated NCs emit linearly polarized light along the elongated axis. This was verified by polarization experiments on single colloidal CdSe quantum rods.[89] This phenomenon was attributed to both a combination of a classical electromagnetic dielectric effect and a quantum mechanical contribution leading to a band-edge electronic transition that is polarized along the rod axis.[91,92]

The mixed 0D/1D CdSe/CdS dot-in-rod heterostructure discussed above is an interesting case in this context.[83] It showed linearly polarized emission along its elongated axis, which was directly confirmed by using combined optical and atomic force microscopy to correlate the emission polarization with the orientation of single seeded nanorods (Figure 6c).[86] As discussed above, this mixed 0D-1D system allows precise control of the emission wavelength, and also retains the linear polarization functionality inherited from the 1D part. A CdSe/CdS rod-in-rod heterostructure manifesting a 1D-in-1D system exhibited an even higher degree of linearly polarized emission.[38]

CdSe nanoplatelets[93] (NPLs) with a 2D electronic structure and their resulting core-shell heterostructures[94] are other potential nanomaterials with anisotropic optical properties. The nanoplatelets provide the highest color purity as a result of the possibility to control their thickness down to the atomic level,[95] and were also confirmed to show moderate emission polarization.[96]

The versatile selection of anisotropic SCNCs with mixed dimensionality presented so far can be exploited in QD LCD technology (see Section 5) by replacing spherical QDs with nanostructures that emit linearly polarized light with high color purity. In addition to the benefits of LCDs with SCNC backlighting, such devices may efficiently transmit light through the first polarizer filter, thereby possibly leading to another improvement in efficiency in QD LCD displays. Anistropic SCNCs also affect the emission directionality, which has relevance for light output from displays.

### 3.5. Fluorescence "Blinking" and its Suppression

SCNCs often exhibit intermittency in their emission, which is called PL "blinking". Until single nanocrystal measurements were conducted, the blinking nature of the SCNCs was hidden in the stable PL of the ensemble. This effect was first noted by Nirmal et al. in studies on isolated CdSe NCs.[97] Blinking is a significant phenomenon in both photo- and electroluminescent display applications of SCNCs. As confirmed by using correlated AFM/optical imaging of single QDs,[98] bright SCNCs can have near-unity quantum yields, thus, blinking remains the last inherent obstacle from achieving near unity QY emitters for efficient display applications.

A typical time-dependent PL intensity trace from single SCNCs (Figure 7a) is recognized by a binary time-dependent intensity, where the emission flickers between "on" and "off" states in a random fashion. However, initial studies by Kuno et al. over a dynamic range of timescales yielded a very surprising result.[99] Neither the "on" nor "off" blinking times follow single exponentials, but rather a power-law distribution over some eight to nine orders of magnitude in the blinking probability density. This implies that there is no characteristic time for the blinking process. This fact has made the photophysics of blinking an intriguing subject that is still under debate.[100, 101]

Currently, the most common explanation for blinking is the charging/discharging model (type A blinking), which involves a nonradiative Auger process (Figure 7a).[102] During the "on" state, each excitation of an electron-hole pair is followed by a radiative recombination process. However, upon charging of the SCNC core, the PL intensity switches from "on" to "off". Charging can happen by several mechanisms,[100] most commonly by trapping of an electron (or hole) in a surface or interfacial trap state. In the presence of excess charge, a nonradiative Auger recombination process occurs efficiently in the small confined NC. In this process, the additional exciton energy is efficiently transferred to the extra electron or hole instead of being emitted as a photon. Release from the trapped state leads to return of the excess carrier to the nanocrystal and its neutralization, thereby restoring the "on" state.

Interestingly, the charging/discharging model predicts a shorter lifetime during the "off" state because of the typically much faster Auger process; however, experiments on CdSe/CdS core-shell NCs revealed stable PL lifetimes during the intermittency. The stable lifetime was attributed to hot electrons that were trapped in surface states and then underwent fast nonradiative recombination with the hole before decaying to the band edge. As a result, the radiative recombination dynamics of the band edge continues on their normal fluorescence timescale, but with a much-reduced QY. To explain the PL flickering in this second type of blinking (known as type B), activation and deactivation of some surface trap states were suggested, and were tested by spectro-electrochemistry on single NCs.[102]

Recently, non-blinking NCs started to emerge. Thick-shell SCNCs with non-blinking characteristics were reported for a CdSe core overcoated with a thick CdS shell.[103, 104] Independent of the specific blinking model, the occurrence of "off" periods during PL blinking is associated with a trapping of one of the carriers at the sites located outside the CQD. Non-blinking emission can be achieved by blocking trap sites with thick shells, which prevent the tunneling of charge carriers to surface traps (Figure 7b). However, thick shells also often exhibit independent fluorescence from the shell material, which is undesired.

Another important approach for suppressing Auger recombination was proposed in the theoretical study by Cragg and Efros.[105] Their studies indicated that, in addition to the QD size, the shape of the confinement potential arising from a composition gradient also has a significant effect on the rate of the Auger decay (Figure 7b). The rate of the Auger process depends on the strength of the intraband transition, where the third carrier is excited to a higher-energy state. As a consequence of its large energy ($E_g$ above the band edge), this state is characterized by a rapidly oscillating wave function ($\Psi_f$ in Figure 7), with its Fourier transform narrowly distributed around a high $k$ value in the $k$ space. The Fourier transform of the ground-state wave function, on the other hand, is centered on $k=0$ and its spreading into the $k$ space is dictated by the shape of the confinement potential. For example, its spreading extends to large values of $k$ in the case of an abrupt square-shaped potential, while its spreading is reduced in the case of a smooth potential barrier. As a result, the Auger transition matrix element, which depends on the overlap of the distribution around the *k* values of the initial and final states, is smaller in the case of a more gradual confinement potential.[106]

This theoretical study was verified experimentally for several gradient shell systems, for example, for a $CdSe_xS_{1-x}$ alloy layer with a controlled composition and thickness between the CdSe core and the CdS shell.[107-109] An increased biexciton lifetime was observed that correlates with the reduced Auger rate. The importance of the grading interface was also shown for other dimensionalities, as seen in $CdSe/Cd_{1-x}Zn_xS$-seeded nanorods with a radially graded composition that show bright and highly polarized green emission with minimal blinking.[41] These observations provide direct experimental evidence that, in addition to the size of the quantum dot, its interfacial properties also significantly affect the rate of Auger recombination and can be used as an additional parameter for tuning the optical properties of SCNCs for display applications.

## 4. Patterning and Alignment Methods for Display Applications of SCNCs

### 4.1. Patterning Methods

There is a need in various display applications to pattern a set of red-green-blue-emitting QDs into defined pixels at high resolution. For example, for the 4K ultrahigh-definition standard and a display size of 22.2 inches, the pixel dimensions are on the scale of 100 micrometers, and for smartphone displays the dimensions can be as small as tens of micrometers. The chemical processability inherent to the solution-based SCNCs opens various possibilities for scalable patterning approaches that are applicable to display manufacturing.

In transfer (contact) printing, patterns are replicated by an elastomeric stamp produced through lithographic means (Figure 8a). Initially, a uniform film of CQDs with a particular color is spin cast on donor substrates to yield an ink pad. The CQDs are weakly bound to the surface, and upon pressing a polydimethylsiloxane (PDMS) stamp to the pad, the CQDs adhere to the stamp, and the stamp is quickly peeled back. The relationship between the pressure applied to the stamp and the peeling velocity influences the efficiency of the pickup process. The stamp is then brought into contact with a receiving surface and slowly peeled back, to allow for complete transfer of the CQDs to the receiving substrate. This process is sequentially repeated for the remaining two colors to yield a compact pattern of three separate colors. This was used for the fabrication of a 4-inch full-color active-matrix CQD display with a resolution of 320×240 pixels by using CdSe/CdS/ZnS red-emitting CQDs, and CdSe/CdS green- and blue-emitting CQDs.[110]

The main advantage of this procedure is the ability to pattern relatively large areas simultaneously and, since a "dry" ink is used, this may reduce pixel cross-talk. However, transfer printing requires the preparation of an etched stamp by lithographic means and it has an inherent serial nature in its ability to pattern only one colored CQD layer at a time. This technique also suffers from deteriorating stamping quality as the resolution of the pixel increases (Figure 8a). An intaglio transfer printing technique was developed to overcome this disadvantage, and is claimed to demonstrate improved CQD transfer efficiencies for smaller features.[111] The basic process is similar to that of conventional transfer printing; however, instead of etching the desired pattern on the stamp, the stamp remains featureless in the intaglio method, and following the ink loading stage, the stamp is pressed on an engraved substrate. In this stage, a pattern of ink remains on the stamp to match the pattern of the engraved features, while all excess ink is removed (Figure 8a). In this way, pressing the stamp with the remaining ink onto a receiving substrate results in a pattern with defined borders being drawn.

Another technique is the ink-jet printing method. This is a relatively simple and material-conserving technique, with the ability to program mask-free arbitrary patterns. It is also compatible with multiple "ink" materials. Ink-jet printing employs a nozzle head, which is electrically controlled to drop a fixed volume of ink on-demand. Once the drops hit the surface, they spread and dry to form a thin film. The ink-jet technique is well-established to provide precision along with the ability to generate almost any pattern.[112-114] For QLED applications, there can be three adjacent nozzle heads which can eject droplets of the three different colors simultaneously. This substantially simplifies the patterning process as all the colors are deposited in one stage. However, this method suffers from two drawbacks: the first is possible re-dissolution of the printed layers during the processing stage and the second is the coffee-ring effect, in which a drying pattern of CQDs concentrated at the rim of the droplet is formed.[118]

The first disadvantage can be treated through the ink formulation process. The ink solution must be adjusted to the right concentration, viscosity, and volatility to match the parameters of both the printing technique and the formation of a stacked structure. The second disadvantage was recently addressed by using a formulation solution that demonstrated a low surface tension, which yielded patterns without coffee-ring effects.[119] Recently, a full-color QLED display was fabricated by ink-jet printing (Figure 8b).[115] In this approach, cross-contamination of pixels was avoided by using a hydrophobic pixel-defining layer to confine the QD inks to their pixel areas.

Electrohydrodynamic jet (e-jet) printing was used (Figure 8c) to improve the lateral resolution and control over the thickness of the deposited features. In this method a voltage bias is applied between a substrate and a metal-coated glass capillary. This induces rapid flow of the CQD ink through the nozzle. To gain control over the thickness of the pattern, a sequence of overlaid printing is defined. Thicker patterns do not demonstrate a significant change in the lateral dimensions and provide increased fluorescence.[116]

An additional technique which can be applied to produce pixels on large areas with micrometer-scale resolution and thickness control is the layer-by-layer (LbL) self-assembly approach combined with photolithography (Figure 8d).[117] A photoresist (PR) layer is cross-linked with UV light through a mask and then exposed to $O_2$ plasma to endow the substrate with a negative charge. By consecutively dipping the substrate in solutions of positively charged polyelectrolytes followed by a solution of NCs with hydrophilic negatively charged ligands, a layer of NCs is formed with the thickness controlled by the number of deposition cycles. Immersing the substrate in acetone (lift-off process) then affords a CQD patterned substrate. Since the previously patterned CQDs are robust against additional photolithographic processes, the procedure described above can be repeated for the remaining subpixels. By using this method, a CQD-LED-based display was fabricated.

### 4.2. Alignment Methods

The polarization and light-output characteristics of SCNCs are also highly relevant for displays. For example, in LCD devices within a backlighting panel, an aligned nanorod array can absorb unpolarized light while re-emitting polarized light. The linearly polarized luminescence along with the directional emission from individual nanostructures can result in enhancement of the brightness, compared to the emitter layers with randomly oriented nanostructures. Harnessing the property of polarized emission from an ensemble of particles requires the development of assembly techniques which provide unidirectional alignment of the NCs.[120]

An external electric field can be used to align anisotropic NCs with their elongated axis oriented along the direction of the field lines. This was reported for the self-assembly of 1D CdSe-seeded CdS nanorods on a scale of tens of square micrometers, which demonstrated polarization ratios of 45%.[83] A solution of NCs is drop cast between two electrodes and a voltage is applied while the solution slowly evaporates. The rods orient along the direction of the field lines (Figure 9a,b). A similar approach, but with a slightly different set-up, was used to align CdS rods perpendicular to the substrate with a hexagonal close-packed pattern.[121]

Although both studies demonstrate an effective alignment, their ability to align large areas will require the use of much higher electric fields. In that sense, the technique of mechanical rubbing, widely used for inducing the alignment of liquid-crystal molecules in LCD technology, might offer a more practical solution.[122] It was also implemented on nanorods, where a glass substrate was first spin coated with a solution of CdSe/CdS seeded rods to achieve a thick layer, and then a rubbing cloth was pressed onto the substrate while it rotated.[123] Whereas the fibers of the rubbing cloth pass over the NR film, the NRs align along the direction of the trenches in a plough-like manner, thereby resulting in a unidirectional assembly of the rods (Figure 9c).

The use of mechanical force to align anisotropic NCs embedded in SCNC-polymer composite films has also been reported. For example, the alignment of both CdSe/CdS dot-in-rod and rod-in-rod heterostructures, as well as colloidal CdSe nanoplatelets dispersed in a polymer film, has been achieved by mechanically stretching the film to 200-300% of its original size.[96] The aligned NR arrays resulted in a twofold enhancement of the brightness of the composite films compared to the emitter layers with randomly oriented nanostructures.

Another approach for inducing the directed alignment of NCs is to utilize templates with either topographical and/or chemical patterns. Such templates, which most often exhibit submicrometer patterns, are available by lithographic means or by exploiting an inherent pattern of a material (such as carbon nanotubes, highly oriented pyrolytic graphite (HOPG) substrates, or block copolymers) as a platform onto which the NCs can be ordered into aligned arrays.

Block copolymers (BCPs) are of particular interest for the alignment of anisotropic SCNCs. As a consequence of the chemical immiscibility of the different blocks, BCPs tend to undergo microphase separation into periodic nanopatterns. When the BCPs are prepared as a thin film, the pattern can be extended to an area on the $cm^2$ scale. By relying on different intermolecular forces between the NRs and each of the blocks, mixing NRs with BCPs in the solution phase, followed by film preparation and an annealing process can lead to not only the selective placement of the rods in only one phase (i.e. block) but also to the alignment of the rods along the patterns exhibited by the BCP. The alignment of CdSe NRs was achieved by using this method (Figure 9d).[124, 126, 127] In addition, changing the ratio between the size of the NRs and the width of the BCP domain enabled control over the perpendicular or parallel orientation of the rods relative to the BCP domain.

Alignment along an interface is another option for aligning NCs.[128] For this purpose, silicon substrates were vertically dipped in a vial containing a concentrated solution of NRs. By gently pulling out the substrate from the vial, slow evaporation of the solution along the solution-substrate interface yielded directional, smectic-type ordering of the NRs. The use of a substrate patterned with electrodes enabled alignment of the NRs in ribbons running parallel to the surface, vertically to the electrodes surface (Figure 9e).[125]

Photo-lithographically patterned silicon nitride surfaces were also used to align needle-like CdSe/CdS superparticles through capillary forces.[129] The aligned superparticles were then transferred into macroscopic and free-standing PDMS films. These superparticle-PDMS composite thin films are highly transparent and exhibit strong linearly polarized PL, with an emission polarization ratio even larger than the typical emission polarization ratio of individual single CdSe/CdS nanorods. This enhancement was attributed to dielectric effects and to collective electric-dipole coupling effects. In addition, the potential applicability of the composite films was demonstrated by their utilization as energy down-conversion phosphors to create polarized light-emitting diodes (LEDs).

## 5. SCNCs as a Down-Converting Backlight for Color Enrichment in Displays

The first commercial application of CQDs in displays was for color enrichment as part of the display back-light unit (BLU). Since 2013, LCD tablets, monitors, and televisions using CQD-based BLU technology are at the technological forefront for display manufacturers. This type of application of CQDs does not require patterning and has been a good in-road for their implementation in displays, as they can be integrated as a drop-in solution that is fully compatible with the highly complex and advanced technologies for LCD fabrication.

The emission efficiency and color quality of SCNCs, which are critical characteristics for their application in displays, have made them an alternative for current down-converting materials in BLUs. Much is being made to develop the next-generation of emitting materials that can cover the full-color gamut according to the standard requirements. Essentially all imaging-based applications need a specific well-defined color gamut to accurately reproduce the colors in the image content. What makes a color gamut an obligatory standard is the growing content created specifically for that color gamut, and thus display manufacturers strive to include that standard in their products. Over the years this has given rise to many different color gamut standards, based on what the existing displays at that specific time could produce. Both the displays and content have evolved together over time.

Looking ahead, Rec. 2020 is an International Telecommunication Union (ITU) Recommendation, first introduced in 2012, that sets out the standards for UHDTV (ultrahigh-definition television).[130] Included in these standards is the Rec. 2020 Color Space, which is an RGB color space that has a color gamut that is wider than all other RGB color spaces. SCNCs have inherently important advantages for addressing this demand. One of their most important strengths is the narrow band emission and efficiency, which extends the color gamut of the displays significantly.

Figure 10a illustrates a typical structure of an LCD display with SCNCs. It contains a blue LED source, a diffusing plate, an SCNC film, a liquid-crystal module (LCM) unit, polarizers, and color filters. The BLU provides a uniform, white sheet of light behind the LCM of the display. The LCM contains millions of pixels, with each of them split into subpixels, with green, red, and blue filters. The color filter separates its component color from the white light of the BLU. By controlling the amount of light passing through each of the color filters, any color that can be composed of a combination of red, green, and blue can be displayed at each pixel location.

Today, white light is typically generated in LCD displays from YAG-rare-earth-based white LEDs (i.e. an yttrium-aluminum-garnet with rare-earth phosphor pumped by a blue GaN LED source). These sources produce a narrow peak in the blue region but with a broad spectrum throughout the yellow and red regions (Figure 10b). When this light is filtered in the RGB color filters at each of the subpixels, the brightness is compromised. Therefore, for the sake of efficiency, an ideal light source should generate narrow peaks in the red, green, and blue regions defined by the color filters with minimal intensity in between, as this light will eventually be attenuated in the color filters.

Unlike phosphor technologies such as YAG, which emit with a fixed spectrum, SCNCs can be synthesized to convert light into a vast color range in the visible spectrum. Pumped with a blue source, such as the GaN LED, they can be engineered to emit at any desired wavelength with very high efficiency (quantum yield near unity) and with a very narrow spectral width (Figure 10b) of only 30-40 nm FWHM (full width at half maximum). Thus, the ability to tune and match the backlight spectrum to the color filters forms the basis of the CQD BLU technology. This yields displays that are brighter, more efficient, and produce lively colors.

Several studies have addressed the required FWHM of the NC emission to be able to meet different color standards. Erdem and Demir concluded that the emission FWHM of the QDs should be at most 50 nm to cover 100% of the NTSC (National Television System Committee) color gamut,[131] otherwise the green end of the NTSC triangle cannot be included. Additionally, the blue emission should be generated with a source having an FWHM of no more than 70 nm. Zhu et al. showed that the FWHM of the QD emission can affect the display performance.[132] As the linewidth of green CQDs and red CQDs increases from 10 nm to 50 nm, both the color gamut and efficiency are reduced. Therefore, a narrower SCNC emission is needed to reach both a larger color gamut and high efficiencies.

SCNCs have another inherent potential advantage particularly for LCD display technology. As discussed in Section 3.4, precise control over the dimensionality and shape of the SCNCs makes it possible to synthesize SCNCs with anisotropic shapes such as rods, dot-in-rods,[82, 83] platelets,[93] or dumbbells.[133] SCNCs with elongated shapes emit polarized light.[134] In the case of LCD display panels, this may turn into an advantage. Whereas standard phosphors or spherical CQDs emit unpolarized light, which in principle leads to a 50% attenuation at the first polarizer of the LCM, elongated SCNCs can be aligned to emit a certain polarization which corresponds to the polarizer orientation, with a potential advantage for higher efficiency (Figure 10a).

To realize and employ the great optical performance of SCNCs in displays, they have to be integrated into the optimal position in the BLU to demonstrate a stable operation that meets the stringent stability standards of state-of-the-art displays. Three typical packing structures to incorporate SCNCs into BLU are on-chip, on-edge, and on-surface types.[135]

These different architectures can be realized in practice by taking advantage of the flexibility in tuning the nanocrystal surface ligands such that they will match the specific required matrices in each case. In the on-chip geometry, the SCNC material is coated onto the emissive LED surface within the LED package. The advantage of this method is that a minimum area is coated with SCNC material, thus minimizing the quantity of material consumed. However, the SCNC material must withstand extreme conditions, as typical LEDs operate at junction temperatures of 85-120°C. Moreover, the SCNC material is exposed to a high flux of blue light, and this also leads to an increased temperature as a result of energy loss through down-conversion that translates into heat. Thus, an SCNC material on-chip might need to operate at about 150°C.[136] Therefore, efforts are invested in stabilizing SCNCs for this demanding task.

In the on-edge solutions, the SCNCs are placed adjacent to the blue LEDs on the edge of the waveguide. The coating area is much larger than that found in the on-chip scenario, but the operating conditions are also less harsh. This approach was realized by embedding the SCNCs into glass tubes. From a practical viewpoint, it is a rather complex task to integrate the glass tubes into the display, and since they are fragile components, their utilization for mobile devices is challenging.

In the on-surface (film) solutions,[137] the SCNCs are embedded in a sheet as part illuminated by the blue LEDs forming the BLU. As a result, the demands on the material are much reduced, as the operating temperature of the SCNCs in this configuration may be very close to room temperature and the optical flux through the NC material is low. This offers a seamless approach for display integration and it is also applicable to mobile displays.

LCD displays with SCNC films are, therefore, a rising power in state-of-the-art displays. They lead to displays with a vast color range, color purity, efficiency, and stability (Figure 10c).

## 6. Electroluminescent Displays with Colloidal SCNCs

EL-driven displays incorporating colloidal SCNCs (EL QD-LEDs) have also attracted significant attention. The advantage over the color-enrichment applications of SCNCs materials is the fact that electroluminescent devices may remove the need for color filters and polarizers if pixelated LEDs are to be utilized. Therefore, they may offer overall high efficiencies if stability issues are addressed and extraction efficiencies are increased. Additionally, EL displays can offer the highest contrast by providing true black (off) pixels.

EL is the phenomenon of light emission from a material excited by electric current. EL in displays of SCNCs is the result of radiative recombination of charge carriers (electrons and holes) that are injected into an emissive layer from contact electrodes, often through charge transport layers (CTLs) having a suitable band alignment.

Similar to conventional light-emitting diodes (LEDs), EL QD-LEDs typically have a p-i-n structure, which consists an anode, an HTL, a QD emissive layer (EML), an ETL, and a cathode (Figure 11a,b). Under forward bias, electrons and holes are injected from opposite electrodes and delivered by CTLs into the QD EML, in which the injected carriers recombine and generate photons. Photons produced in the EML have an emission energy typical for the SCNC band gap.

Over the years, EL QD-LEDs have been developed extensively and tested with various architectures. These architectures can be categorized in general by the CTLs embedded within them. The first device was comprised only of an organic QD bilayer structure, where the QD layer served both as the ETL and EML, and the PPV polymer served as an HTL.[11, 14] However, this device exhibited a very low EQE (external quantum efficiency) of 0.001-0.01%, poor brightness (ca. 100 cd·m$^{-2}$), and unwanted parasitic polymer-related emission in addition to the emission of the NCs.

To resolve these issues, the next generation of devices used architectures in which the CQDs serve only as light emitters. In this architecture, a single monolayer of CQDs is sandwiched between an organic ETL and an organic HTL.[138] The incorporation of another ETL helps to repress the electron injection into the EML. In addition, this layer also serves as the HBL (hole blocking layer), serving as a barrier to the flow of holes outside the EML. These devices show higher EQE, and no parasitic polymer emission is observed.

However, even after embedding the ETL between the EML and the cathode, the all-organic CTL architecture still has an inherent shortcoming. The LUMO energy level of traditionally used organic layers is higher than the conduction band of Cd-based CQDs, while the holes experience an energy barrier between the HTL and the EML. This leads to excess electron injection into the EML, thereby resulting in nonradiative Auger recombination.

In addition to the efforts on the all-organic architecture, new all-inorganic architectures have been developed, motivated by their long-term stability and their preferred electric conductivity. Inorganic layers can sustain a current density of up to 4000 mA·cm$^{-2}$, whereas all-organic layers so far have a maximum current density of about 1000 mA·cm$^{-2}$.[139] The first all-inorganic device was comprised from a CdSe/ZnS CQD monolayer embedded within p- and n-doped GaN CTLs in a type I band alignment.[140] This architecture leads to a better charge balance in the EML together with the advantages of inorganic layers discussed above.

Several other all-inorganic CTLs were tested. These efforts mainly focused on metal oxides serving as both the ETLs (ZnO, SnO$_2$, ZnO:SnO$_2$, and ZnS)[141] and HTLs (NiO).[142] However, the efficiencies of the architectures with all-inorganic CTLs (EQE <0.01% for the p-GaN/QD/n-GaN structures[140] and <0.1% for the NiO/QD/ZnO:SnO$_2$ structures[142]) were lower than those of devices with all-organic injection layers. This was attributed mainly to the harsh deposition conditions of the inorganic CTL which can damage the EML of the SCNCs[140] as well as to a nonradiative energy transfer to the heavily doped inorganic films.[141]

Significant progress in the field of EL QD-LEDs was achieved by moving to hybrid CTL architectures which combine an inorganic ETL and an organic HTL. The main important advance in hybrid CTLs was made by utilizing solution-processible ZnO nanoparticles as the ETL.[143] In this method, the conductivity robustness of an inorganic CTL is achieved without the harmful consequences of harsh fabrication conditions. Furthermore, the band-edge positions in a ZnO nanoparticle provide a better charge balance in the EML and also prevent hole leakage from the QD EML. Qian et al. have also used a similar approach to fabricate red, green, and blue solution-processed QD-LEDs with brightness values of 31,000 cd·m$^{-2}$, 68,000 cd·m$^{-2}$, and 4,200 cd·m$^{-2}$ for the red, green, and blue devices, respectively.[144] Kwak et al. also used solution-processed ZnO nanoparticles as the ETL, but this time with an inverted architecture and by optimizing the energy levels with the organic hole transport layer, to demonstrate highly bright red, green, and blue QLEDs showing an EQE of 7.3%, 5.8%, and 1.7%, respectively.[145] Such "hybrid CTL" devices currently exhibit the highest efficiency and brightness.[146]

A full-color 4-inch QD-LED display (Figure 11c) has also been fabricated by using similar "hybrid CTL" structures.[110] The pixels were patterned by a microcontact printing method with a resolution of up to 1000 pixels per inch (see Section 4). This display was the first demonstration of a full-color EL QD-LED display demonstrating the applicability of such a concept of using SCNCs in display applications.

Aside from efficient transport layers, the SCNC surface ligands should also be engineered to improve the charge transport in electroluminescent displays. The general strategy to improve the transport is to exchange the ligands with shorter and often conjugated ligands.[147, 148] The right spacing between adjacent NCs is needed to obtain a delicate balance between the charge injection and undesired exciton dissociation. An optimum distance was indeed demonstrated by tuning the distance between adjacent PbS quantum dots through modifying the lengths of the linker molecules from three to eight $CH_2$ groups.[149] In addition, an increase in the EQE by replacing the oleic acid ligands of the as-synthesized SCNCs with shorter 1-octanethiol ligands was reported for violet-blue-emitting ZnCdS/ZnS graded core-shell QDs.[150]

Efficient electron and hole injection as well as charge balance at the CQD active layer are important design criteria for ensuring high-performance LEDs for electroluminescent displays. However, even after bringing charge carriers into the CQD layer and forming excitons on the QDs, the last step in the generation of light in EL CQD-LEDs brings with it inherent challenges that must be addressed for their future commercialization.

SCNCs can today reach near unity QY in solution. However, SCNCs serving as the EML of EL CQD-LED devices operate under conditions which make it hard to emit. The difference between the electron and hole injection efficiencies is likely to create charge imbalance and, hence, a very efficient Auger process. Furthermore, the EML experiences an electric field on the order of 1 MV·cm$^{-1}$, which can lead to a spatial separation between the electron and hole wave functions that reduces the radiative rate of the SCNCs.[151]

Recent studies on the chemistry of SCNCs have exploited thick-shell CdSe/CdS QDs to improve the QY by eliminating blinking. In this system, the hole wave function remains confined at the CdSe core, while the electron wave function extends into the shell.[152,153] However, the electric field across the EML will decrease the luminescence QY by inducing charge separation.[154] On the other hand, using CdSe/ZnS CQDs with a distinct type I band alignment that can help to reduce the electron-hole separation even under an electric field will clearly fail to prevent nonradiative Auger processes, especially in EL QD-LEDs, which operate with a continuous charge imbalance.

Based on these considerations, it seems that SCNCs with an alloy composition, such as $Zn_xCd_{1-x}S$, $Zn_xCd_{1-x}Se$, or $CdSe_xS_{1-x}/Zn_xCd_{1-x}S$ dot-in-rod structures,[41] which offer confined potential against the electric field and a reduction of Auger nonradiative recombination by smoothing the confinement potential, would be the ideal choice for an emitter in an EL QD-LED. Indeed, relatively high efficiencies in QD-LEDs have been reported with multilayered alloyed structures such as $Zn_xCd_{1-x}Se/ZnS$, $Zn_xCd_{1-x}Se/Zn_xCd_{1-x}S$, and $Zn_xCd_{1-x}Se$.[139,142,155] Recently, Yang et al. reported a full series of blue, green, and red QD-LEDs, all with EQE values over 10% by using graded-shell $Cd_{1-x}Zn_xSe_{1-y}S_y$ as well as an ETL of ZnO nanoparticles in "hybrid CTL" architectures.[146]

Another challenge in electroluminescent QD-LEDs is to couple the light outside the layer structure. The fraction of emitted light which escapes from the surface of a device is determined by the difference in the refractive index of the organic and inorganic layers of the LED, which defines an "escape cone" for each interface, outside of which light will remain trapped. The light outside of the "escape cone" propagates within the plane of the device. A fraction of the light can return into the "escape cone" through a scattering event. A larger fraction is eventually lost to absorption by the CTL, reabsorbed in the EML, or attenuated by exciting surface plasmon modes of the metal electrode.[156] All together, the upper limit for the out-coupling efficiency is on the order of about 20%.

The use of elongated-shape SCNCs can improve the out-coupling problem.[157] As discussed in Section 4, elongated-shape SCNCs favor emission along the elongated axis, and the elongated structure results in preferential flat deposition on the substrate. Therefore, the emission direction couples to only a very small extent with the wave-guiding modes or the plasmonic modes of the electrode. This is another manifestation of how diversity in the size, shape, and composition of SCNCs can be engineered to improve the efficiency of QD-LEDs.

Despite the present low efficiencies, stability challenges, and manufacturing issues of electroluminescent QD-LEDs, this type of display has an inherent benefit: the per-pixel control, which leads to improved contrast. Per-pixel control can help to reach the next level of picture quality, with lifelike contrast ratios. This is the main reason for the on-going research into the optimization of QD-LEDs that eventually may emerge as important players in the display market.

## 7. Prospects and Challenges for SCNCs in the Field of Displays

### 7.1. New Functionalities and New Modes of Operation

The discussion above on the challenges associated with designing SCNCs that exhibit high QY in the presence of both charge and an electric field highlights the interest in new systems that shift away from charge injection into the NCs, towards device architectures that are not p-n junctions and that operate through electrically controlled SCNC emission in new ways. Since the early studies of Rothenberg et al. on the optical properties of single nanorods under electric field bias,[158] it has been known that the electric field influences the emission intensity through the quantum-confined Stark effect.[159] This is highly pronounced in nanorods aligned along the direction of the electric field, with electrons and holes being pulled in opposite directions towards the ends of the rods. The result is that the overlap of their wave functions is reduced, and the optical transition decreases in intensity (because of a decrease in the oscillator strength). The process depends strongly on the mutual orientations of the electric field and the long axis of the rod. If they are parallel, the effect is maximized, whereas if they are perpendicular, the influence of the electric field is negligible. Further studies by Muller et al. on CdSe/CdS dot-in-rod heterostructures with quasi type II band alignment showed that the wave function distributions of the electrons and holes were modified greatly from a configuration with a large overlap to a spatial separation where the holes were localized in the core and the electrons forced to the opposite tip by the electric field.[160] In this case, the emission intensity was also largely reduced due to the small spatial overlap of the electron and hole wave functions (Figure 12).

These experiments pave the way to a possible new mode of operation for QD-LEDs controlled by an electric field. In this mode of operation, the excitons are created in a subpixel containing a large number of aligned SC nanorods emitting red, green, or blue light. Each subpixel is found between electrodes and their optical excitation from the light coming from the BLU, can modulate the light intensity of each color, thereby creating any desirable color composed from red, green, and blue. This mode of operation utilizes the high QY of SCNCs excited optically, but at the same time dispenses with the need for the charge injection EL architecture as well as the liquid-crystal unit in LCD displays.

Besides new modes of operation, new functionalities are becoming apparent in QD-LED displays containing SCNCs. Recently, a novel study by Oh et al. presented DHNRs comprised of CdSe tips at the end of a CdS rod, with the CdSe tips coated with a ZnSe shell. These DHNRs acted as both charge-separation and recombination centers.[12] These particles can simultaneously enable efficient photocurrent generation and EL by changing from a reverse to a forward bias (Figure 13). These devices exhibit high efficiencies at display-relevant brightness with an external quantum efficiency of 8.0% at 1000 cd·m$^{-2}$ under 2.5 V bias.[12]

This study opens the way to a new type of display panel that moves from just displaying information towards interactive devices. Possible ways to use the multifunctionality of these devices could be in cellphones and other devices which can be controlled through touchless gestures or automatically through brightness control (Figure 13). In addition, such devices can charge themselves, thus paving the way to a situation where a significant amount of the display's power comes from the surroundings. In addition to interacting with users and their environment, DHNR LED displays can interact with each other and form large parallel communication arrays, thereby leading to optical communication between devices.

### 7.2. Heavy-Metal-Free SCNCs

One of the major drawbacks in QD-LED technology incorporating SCNCs is the toxicity problem. Cd-based II-VI NCs are currently the most well-developed and optimized system; they offer highly monodisperse NCs with near unity QY and well-developed shape control that can cover the entire visible spectrum. However, regulations by governments and organizations restricting the amount of heavy metals form an obstacle to Cd-based devices.[161] Thus, developing nontoxic SCNCs that have a similar performance as the Cd-containing systems is becoming an increasingly active area of research.

One alternative in the II-VI group is the Zn-chalcogenide-based SC (Zn/S, Se, Te) NCs. Monodisperse core-shell ZnSe/ZnS QDs with narrow emission peaks (FWHM 15-18 nm) and tunable blue emission have, for example, been developed.[162,163] As a consequence of the relatively greater band gap of Zn-based NCs compared to Cd-based NCs, the emission wavelength is in the blue region of the spectrum. However, the use of type II Zn-based NCs can span a broader range of green to yellow colors. For example, type II ZnTe/ZnSe (core/shell) NCs were demonstrated to emit from cyan to amber.[164]

More recently, Ji et al. presented ZnSe/ZnTe NDBs, in which ZnSe tips were selectively grown on the tips of ZnTe rods.[87] These NDBs can be tuned to emit between about 500-585 nm by changing the size of the ZnSe tip. This example also shows the shape control that was developed in Zn-based NCs, despite their more symmetric zinc blende crystal phase. Starting from 1D Zn-chalcogenide-based nanowires, rods with controllable aspect ratios were synthesized using a ripening process through thermodynamically driven material diffusion.[165]

NCs with mixed dimensionalities such as dot-in-rod and platelets were also obtained. ZnSe(core)/ZnS(shell) nanorods achieved by cation exchange from CdSe(core)/CdS(shell) nanorods were presented by Li et al. that showed relatively narrow line widths (ca. 75 meV).[166] ZnSe/ZnS core-shell platelets were also demonstrated by Bouet et al. by cation exchange.[167] However, despite the progress achieved in Zn-based NCs, only relatively low quantum efficiencies as a result of defect-related emissions have so far been reported. Therefore, the facilitation of efficient Zn-chalcogenide-based NCs in QD-LED displays remains a challenge.

Another promising heavy-metal free NC is InP. It is a III-V SC ($E_g$=1.35 eV) with an exciton Bohr radius of about 10 nm (compared to ca. 5 nm for CdSe) that provides large quantum confinement effects that can be used to span the visible spectrum through size control. Despite being part of the III-V group, which in general need high-temperature synthesis because of their more covalent-type bonding,[168] the groups of Bataglia,[48] Xie,[169] and Yang[170] showed that high-quality InP and InP/ZnS core/shell NCs could be synthesized at relatively low temperatures. The InP/ZnS core/shell NCs can cover the visible spectrum with a QY above 60% and a FWHM as narrow as 38 nm, which is close to that of Cd-based NCs. More recently, a new approach to form InP-based QDs was introduced that is based on aminophosphines as the phosphorus precursor.[171,172] This method enables a variety of InP-based core/shell QDs to be generated. Shape control of III-V CQDs was also achieved by Kan et al. by using metal nanoparticles to direct and catalyze one-dimensional growth.[18] An alternative to the hot-injection synthesis of InP is to attempt to use cation exchange.[173-175] This synthetic route also has the advantage of preserving the original NC shape, thus paving the way towards shape control in these zinc blende NCs.

InP/ZnSeS QD-LEDs that emit between 500 and 520 nm through direct charge carrier injection have also been demonstrated that show an EQE of 3.5 % and a maximum brightness of 3900 cd·m$^{-2}$.[176] This example illustrates the promise of InP as a Cd-free QD-LED emitter. However, efficient blue or red QD-LEDs containing InP NCs have not yet been reported.

Multinary compounds, such as ternary I-III-VI$_2$ chalcogenides, are other candidates for heavy-metal-free NCS. Among these ternary I-III-VI$_2$ compounds, CuIn(S,Se)$_2$ is a candidate for emission in the visible spectrum. It can be synthesized by using a route similar to the cation-exchange approach used to produce III-V NCs starting from $Cu_2S$.[177] Since the size and shape of the NCs remain unchanged, it should again enable shape control, as already exists for $Cu_2S$.[178] However, the complex crystal structures lead to high-density sub-band-gap states being formed that cause the emission to become broader. $CuInS_2$ NCs have also been successfully incorporated into direct-injection LEDs, albeit with modest results.[179,180]

Inorganic perovskite NCs of $CsPbX_3$ (X=Cl, Br, I) are another example of ternary compounds that can serve as possible candidates for future display applications, but they also contain the undesired heavy metal Pb. These materials can be synthesized in a variety of sizes and shapes, with facile compositional tuning.[181-186] Inorganic perovskite NCs demonstrate high chemical stability, especially for blue- and green-emitting perovskites. Their high QY values (ca. 70-90%) and narrow PL spectra (FWHM ranging from ca. 12-40 nm) that span the visible spectrum allows for a wide color gamut representation, and their demonstration as white LED color converters has already been reported.[187,188] These materials also show self-healing, with the structure of the perovskite NCs remaining intact even upon leaching out of metal atoms or diffusion of surfactant atoms to the surface.[184,189] However, the presence of lead in these perovskites presents a problem in terms of toxicity. Alternative lead-free halide perovskites have also been explored, such as $Cs_2SnI_6$,[190] $CsGeI_3$,[191] and even halide double perovskites;[192] however, the first still suffer from instability problems, while the last are still being explored.

Despite the above drawbacks, the performance of QLEDs based on heavy-metal-free emitters is improving. The current limitation is largely due to the undeveloped synthetic chemistry of the heavy-metal-free NCs. However, future developments in SCNC synthesis are expected to close the current gap and lead to the further development of heavy-metal-free SCNC-based displays.

**7.3. Outlook**

The above-told story of the emergence of SCNCs as new materials for current and next-generation displays is a tour-de-force demonstration of the importance of basic research into new nanomaterials to yield important advantages in the highly advanced and demanding field of displays. A combination of innovative well-controlled synthesis, development of suitable chemical processing methods, accompanied by achieving a deep understanding of the effects of the SCNC characteristics on their optoelectronic performance were imperative to reach this point. The ability to upscale the complex synthesis in an economically viable manner has been realized, and the SCNCs can meet the stringent stability requirements of commercial displays. The development of nontoxic SCNCs for next-generation displays is on its way. We foresee more elaborate display technologies based on SCNCs also emerging. This provides an optimistic perspective for the potential utilization of SCNCs in additional areas, including biomedical applications, sensing, solar energy conversion, and printed electronics.

**Acknowledgements**

U.B. wishes to thank the generations of students, postdoctoral researchers, and co-workers who worked with him over the years on research into SCNCs, and the employees of Qlight Nanotech who participated in the development of SCNCs for display applications. This project received funding in part from the European Research Council (ERC) under the European Union's Horizon 2020 research and innovation programme (grant agreement No [741767]), as well as from the Israel Science Foundation (ISF, Grant No. 811/13). U.B. holds the Alfred & Erica Larisch Memorial Chair.

**Conflict of interest**

The authors declare no conflict of interest.

**Author Contributions**

Y.P. Writing – original draft: Equal; Writing – review & editing: Equal.

M.O. Writing – original draft: Equal; Writing – review & editing: Equal.

U.B. Writing – original draft: Equal; Writing – review & editing: Equal.

[1] A. P. Alivisatos, *J. Phys. Chem.* **1996**, *100*, 13226-13239.

[2] H. Weller, *Angew. Chem. Int. Ed. Engl.* **1993**, *32*, 41-53; *Angew. Chem.* **1993**, *105*, 43-55.

[3] M. G. Bawendi, M. L. Steigerwald, L. E. Brus, *Annu. Rev. Phys. Chem.* **1990**, *41*, 477-496.

[4] R. Krahne, G. Morello, A. Figuerola, C. George, S. Deka, L. Manna, *Phys. Rep.* **2011**, *501*, 75-221.

[5] J. M. Pietryga, Y. S. Park, J. Lim, A. F. Fidler, W. K. Bae, S. Brovelli, V. I. Klimov, *Chem. Rev.* **2016**, *116*, 10513-10622.

[6] A. Sitt, I. Hadar, U. Banin, *Nano Today* **2013**, *8*, 494-513.

[7] J. Y. Kim, O. Voznyy, D. Zhitomirsky, E. H. Sargent, *Adv. Mater.* **2013**, *25*, 4986-5010.

[8] M. A. Boles, D. Ling, T. Hyeon, D. V. Talapin, *Nat. Mater.* **2016**, *15*, 141-153.

[9] A. L. Efros, A. L. Efros, *Sov. Phys. Semicond.* **1982**, *16*, 772-775.

[10] L. E. Brus, *J. Chem. Phys.* **1983**, *79*, 5566-5571.

[11] V. L. Colvin, M. C. Schlamp, A. P. Alivisatos, *Nature* **1994**, *370*, 354-357.

[12] N. Oh et al., *Science* **2017**, *355*, 616-619.

[13] B. Mashford, M. Stevenson, Z. Popovic, C. Hamilton, Z. Zhou, C. Breen, J. Steckel, V. Bulovic, M. Bawendi, S. Coe-Sullivan, P. T. Kazlas, *Nat. Photonics* **2013**, *7*, 407-412.

[14] B. O. Dabbousi, M. G. Bawendi, O. Onitsuka, M. F. Rubner, *Appl. Phys. Lett.* **1995**, *66*, 1316-1318.

[15] T. Erdem, H. V. Demir, *Nanophotonics* **2016**, *5*, 74-95.

[16] V. Wood, V. Bulović, *Nano Rev.* **2010**, *1*, 5202.

[17] X. Peng, L. Manna, W. Yang, J. Wickham, E. Scher, A. Kadavanich, A. P. Alivisatos, *Nature* **2000**, *404*, 59-61.

[18] S. Kan, A. Aharoni, T. Mokari, U. Banin, *Faraday Discuss.* **2004**, *125*, 23-38.

[19] B. Dubertret, T. Heine, M. Terrones, *Acc. Chem. Res.* **2015**, *48*, 1-2.

[20] J. S. Son, J. H. Yu, S. G. Kwon, J. Lee, J. Joo, T. Hyeon, *Adv. Mater.* **2011**, *23*, 3214-3219.

[21] C. Schliehe, B. H. Juarez, M. Pelletier, S. Jander, D. Greshnykh, M. Nagel, A. Meyer, S. Foerster, A. Kornowski, C. Klinke, H. Weller, *Science* **2010**, *329*, 550-553.

[22] A. M. Smith, H. Duan, A. M. Mohs, S. Nie, *Adv. Drug Delivery Rev.* **2008**, *60*, 1226-1240.

[23] Z. Tang, Y. Wang, N. A. Kotov, *Langmuir* **2002**, *18*, 7035-7040.

[24] T. Pellegrino, S. Kudera, T. Liedl, A. M. Javier, L. Manna, W. J. Parak, *Small* **2004**, *1*, 48-63.

[25] R. A. Sperling, W. J. Parak, *Philos. Trans. R. Soc. A* **2010**, *368*, 1333-1383.

[26] D. A. Hines, P. V. Kamat, *ACS Appl. Mater. Interfaces* **2014**, *6*, 3041-3057.

[27] R. Freeman, I. Willner, *Chem. Soc. Rev.* **2012**, *41*, 4067.

[28] I. L. Medintz, H. T. Uyeda, E. R. Goldman, H. Mattoussi, *Nat. Mater.* **2005**, *4*, 435-446.

[29] A. Fu, W. Gu, C. Larabell, A. P. Alivisatos, *Curr. Opin. Neurobiol.* **2005**, *15*, 568-575.

[30] P. Alivisatos, *Nat. Biotechnol.* **2004**, *22*, 47-52.

[31] I. J. Kramer, E. H. Sargent, *ACS Nano* **2011**, *5*, 8506-8514.

[32] A. Nozik, *Phys. E* **2002**, *14*, 115-120.

[33] V. I. Klimov, A. A. Mikhailovsky, S. Xu, A. Malko, J. A. Hollingsworth, C. A. Leatherdale, H. Eisler, M. G. Bawendi, *Science* **2000**, *290*, 314-317.

[34] M. Kazes, D. Y. Lewis, Y. Ebenstein, T. Mokari, U. Banin, *Adv. Mater.* **2002**, *14*, 317-321.

[35] Y. Ben-Shahar, U. Banin, *Top. Curr. Chem.* **2016**, *374*, 54.

[36] C. R. Kagan, E. Lifshitz, E. H. Sargent, D. V. Talapin, *Science* **2016**, *353*, aac5523.

[37] R. Xie, U. Kolb, J. Li, T. Basché, A. Mews, *J. Am. Chem. Soc.* **2005**, *127*, 7480-7488.

[38] A. Sitt, A. Salant, G. Menagen, U. Banin, *Nano Lett.* **2011**, *11*, 2054-2060.

[39] A. Robin, E. Lhuillier, X. Z. Xu, S. Ithurria, H. Aubin, A. Ouerghi, B. Dubertret, *Sci. Rep.* **2016**, *6*, 24909.

[40] N. Oh, S. Nam, Y. Zhai, K. Deshpande, P. Trefonas, M. Shim, *Nat. Commun.* **2014**, *5*, 3642.

[41] I. Hadar, J. P. Philbin, Y. E. Panfil, S. Neyshtadt, I. Lieberman, H. Eshet, S. Lazar, E. Rabani, U. Banin, *Nano Lett.* **2017**, *17*, 2524-2531.

[42] J. Park, J. Joo, G. K. Soon, Y. Jang, T. Hyeon, *Angew. Chem. Int. Ed.* **2007**, *46*, 4630-4660; *Angew. Chem.* **2007**, *119*, 4714-4745.

[43] X. Peng, J. Thessing, in *Semicond. Nanocrystals Silic. Nanoparticles* (Eds.: X. Peng, D. M. P. Mingos), Springer, New York, **2005**, pp. 79-119.

[44] C. B. Murray, D. Norris, M. G. Bawendi, *J. Am. Chem. Soc.* **1993**, *115*, 8706-8715.

[45] W. W. Yu, X. Peng, *Angew. Chem. Int. Ed.* **2002**, *41*, 2368-2371; *Angew. Chem.* **2002**, *114*, 2474-2477.

[46] X. Peng, J. Wickham, A. P. Alivisatos, *J. Am. Chem. Soc.* **1998**, *120*, 5343-5344.

[47] C. De Mello Donegá, P. Liljeroth, D. Vanmaekelbergh, *Small* **2005**, *1*, 1152-1162.

[48] D. Battaglia, X. Peng, *Nano Lett.* **2002**, *2*, 1027-1030.

[49] J. Y. Rempel, M. G. Bawendi, K. F. Jensen, *J. Am. Chem. Soc.* **2009**, *131*, 4479-4489.

[50] J. Park, K. An, Y. Hwang, J.-G. Park, H.-J. Noh, J.-Y. Kim, J.-H. Park, N.-M. Hwang, T. Hyeon, *Nat. Mater.* **2004**, *3*, 891-895.

[51] J. Joo, H. Bin Na, T. Yu, J. H. Yu, Y. W. Kim, F. Wu, J. Z. Zhang, T. Hyeon, *J. Am. Chem. Soc.* **2003**, *125*, 11100-11105.

[52] W. W. Yu, J. C. Falkner, C. T. Yavuz, V. L. Colvin, *Chem. Commun.* **2004**, 2306-2307.

[53] C. B. Williamson, D. R. Nevers, T. Hanrath, R. D. Robinson, *J. Am. Chem. Soc.* **2015**, *137*, 15843-15851.

[54] Y. Jun, J. Choi, J. Cheon, *Angew. Chem. Int. Ed.* **2006**, *45*, 3414-3439; *Angew. Chem.* **2006**, *118*, 3492-3517.

[55] S. Kumar, T. Nann, *Small* **2006**, *2*, 316-329.

[56] A. A. Guzelian, U. Banin, A. V. Kadavanich, X. Peng, A. P. Alivisatos, *Appl. Phys. Lett.* **1996**, *69*, 1432-1434.

[57] O. I. Micic, C. J. Curtis, K. M. Jones, J. R. Sprague, A. J. Nozik, *J. Phys. Chem.* **1994**, *98*, 4966-4969.

[58] S. Tamang, C. Lincheneau, Y. Hermans, S. Jeong, P. Reiss, *Chem. Mater.* **2016**, *28*, 2491-2506.

[59] A. L. Efros, M. Rosen, *Annu. Rev. Mater. Sci.* **2000**, *30*, 475-521.

[60] A. M. Smith, S. Nie, *Acc. Chem. Res.* **2010**, *43*, 190-200.

[61] W. E. Buhro, V. L. Colvin, *Nat. Mater.* **2003**, *2*, 138-139.

[62] U. Banin, Y. Cao, D. Katz, O. Millo, *Nature* **1999**, *400*, 542-544.

[63] U. Banin, O. Millo, *Annu. Rev. Phys. Chem.* **2003**, *54*, 463-494.

[64] V. I. Klimov, *J. Phys. Chem. B* **2000**, *104*, 6112-6123.

[65] V. Klimov, D. McBranch, *Phys. Rev. Lett.* **1998**, *80*, 4028-4031.

[66] V. I. Klimov, D. W. McBranch, C. A. Leatherdale, M. G. Bawendi, *Phys. Rev. B* **1999**, *60*, 13740-13749.

[67] R. Ulbricht, E. Hendry, J. Shan, T. F. Heinz, M. Bonn, *Rev. Mod. Phys.* **2011**, *83*, 543-586.

[68] V. I. Klimov, *Annu. Rev. Phys. Chem.* **2007**, *58*, 635-673.

[69] A. R. Kortan, R. Hull, R. L. Opila, M. G. Bawendi, M. L. Steigerwald, P. J. Carroll, L. E. Brus, *J. Am. Chem. Soc.* **1990**, *112*, 1327-1332.

[70] P. Reiss, M. Protière, L. Li, *Small* **2009**, *5*, 154-168.

[71] D. Dorfs, A. Eychmüller in *Semiconductor Nanocrystal Quantum Dots: Synthesis, Assembly, Spectroscopy and Applications*, Springer, Wien, **2008**, pp. 101-117.

[72] C. de Mello Donegá, *Chem. Soc. Rev.* **2011**, *40*, 1512-1546.

[73] L. Carbone, P. D. Cozzoli, *Nano Today* **2010**, *5*, 449-493.

[74] X. Zhong, R. Xie, Y. Zhang, T. Basché, W. Knoll, *Chem. Mater.* **2005**, *17*, 4038-4042.

[75] X. Peng, M. C. Schlamp, A. V. Kadavanich, A. P. Alivisatos, *J. Am. Chem. Soc.* **1997**, *119*, 7019-7029.

[76] M. A. Hines, P. Guyot-Sionnest, *J. Phys. Chem.* **1996**, *100*, 468-471.

[77] S. Kim, B. Fisher, H. J. Eisler, M. Bawendi, *J. Am. Chem. Soc.* **2003**, *125*, 11466-11467.

[78] D. Dorfs, T. Franzi, R. Osovsky, M. Brumer, E. Lifshitz, T. A. Klar, A. Eychmüller, *Small* **2008**, *4*, 1148-1152.

[79] P. Reiss, S. Carayon, J. Bleuse, A. Pron, *Synth. Met.* **2003**, *139,* 649-652.

[80] J. J. Li, Y. A. Wang, W. Guo, J. C. Keay, T. D. Mishima, M. B. Johnson, X. Peng, *J. Am. Chem. Soc.* **2003**, *125*, 12567-12575.

[81] D. Chen, F. Zhao, H. Qi, M. Rutherford, X. Peng, *Chem. Mater.* **2010**, *22*, 1437-1444.

[82] D. V. Talapin, R. Koeppe, S. Götzinger, A. Kornowski, J. M. Lupton, A. L. Rogach, O. Benson, J. Feldmann, H. Weller, *Nano Lett.* **2003**, *3*, 1677-1681.

[83] L. Carbone et al., *Nano Lett.* **2007**, *7*, 2942-2950.

[84] D. V. Talapin, J. H. Nelson, E. V. Shevchenko, S. Aloni, B. Sadtler, A. P. Alivisatos, *Nano Lett.* **2007**, *7*, 2951-2959.

[85] S. Halivni, S. Shemesh, N. Waiskopf, Y. Vinetsky, S. Magdassi, U. Banin, *Nanoscale* **2015**, *7*, 19193-19200.

[86] I. Hadar, G. B. Hitin, A. Sitt, A. Faust, U. Banin, *J. Phys. Chem. Lett.* **2013**, *4*, 502-507.

[87] B. Ji, Y. E. Panfil, U. Banin, *ACS Nano* **2017**, *11*, 7312-7320.

[88] J. Hu, L. Li, W. Yang, L. Manna, L. Wang, A. P. Alivisatos, *Science* **2001**, *292*, 2060-2063.

[89] X. Chen, A. Nazzal, D. Goorskey, M. Xiao, Z. Peng, X. Peng, *Phys. Rev. B* **2001**, *64*, 2-5.

[90] J. Wang, M. S. Gudiksen, X. Duan, Y. Cui, C. M. Lieber, *Science* **2001**, *293*, 1455-1457.

[91] A. Shabaev, A. L. Efros, *Nano Lett.* **2004**, *4*, 1821-1825.

[92] J. Planelles, F. Rajadell, J. I. Climente, *J. Phys. Chem. C* **2016**, *120*, 27724-27730.

[93] S. Ithurria, B. Dubertret, *J. Am. Chem. Soc.* **2008**, *130*, 16504-16505.

[94] B. Mahler, B. Nadal, C. Bouet, G. Patriarche, B. Dubertret, *J. Am. Chem. Soc.* **2012**, *134*, 18591-18598.

[95] F. Zhang, S. Wang, L. Wang, Q. Lin, H. Shen, W. Cao, C. Yang, H. Wang, L. Yu, Z. Du, J. Xue, L. S. Li, *Nanoscale* **2016**, *8*, 12182-12188.

[96] P. D. Cunningham, J. B. Souza, I. Fedin, C. She, B. Lee, D. V. Talapin, *ACS Nano* **2016**, *10*, 5769-5781.

[97] M. Nirmal, B. O. Dabbousi, M. G. Bawendi, J. J. Macklin, J. K. Trautman, T. D. Harris, L. E. Brus, *Nature* **1996**, *383*, 802-804.

[98] Y. Ebenstein, T. Mokari, U. Banin, *Appl. Phys. Lett.* **2002**, *80*, 4033-4035.

[99] M. Kuno, D. P. Fromm, H. F. Hamann, A. Gallagher, D. J. Nesbitt, *J. Chem. Phys.* **2000**, *112*, 3117-3120.

[100] A. L. Efros, D. J. Nesbitt, *Nat. Nanotechnol.* **2016**, *11*, 661-671.

[101] O. Schwartz, D. Oron, *Isr. J. Chem.* **2012**, *52*, 992-1001.

[102] C. Galland, Y. Ghosh, A. Steinbrück, M. Sykora, J. A. Hollingsworth, V. I. Klimov, H. Htoon, *Nature* **2011**, *479*, 203-207.

[103] B. Mahler, P. Spinicelli, S. Buil, X. Quelin, J.-P. Hermier, B. Dubertret, *Nat. Mater.* **2008**, *7*, 659-664.

[104] Y. Chen, J. Vela, H. Htoon, J. L. Casson, D. J. Werder, D. A. Bussian, V. I. Klimov, J. A. Hollingsworth, *J. Am. Chem. Soc.* **2008**, *130*, 5026-5027.

[105] G. E. Cragg, A. L. Efros, *Nano Lett.* **2010**, *10*, 313-317.

[106] J. I. Climente, J. L. Movilla, J. Planelles, *Small* **2012**, *8*, 754-759.

[107] Y. S. Park, W. K. Bae, L. A. Padilha, J. M. Pietryga, V. I. Klimov, *Nano Lett.* **2014**, *14*, 396-402.

[108] K. Boldt, N. Kirkwood, G. A. Beane, P. Mulvaney, *Chem. Mater.* **2013**, *25*, 4731-4738.

[109] J. Vela, H. Htoon, Y. Chen, Y. S. Park, Y. Ghosh, P. M. Goodwin, J. H. Werner, N. P. Wells, J. L. Casson, J. A. Hollingsworth, *J. Biophotonics* **2010**, *3*, 706-717.

[110] T.-H. Kim et al., *Nat. Photonics* **2011**, *5*, 176-182.

[111] M. K. Choi, J. Yang, K. Kang, D. C. Kim, C. Choi, C. Park, S. J. Kim, S. I. Chae, T.-H. Kim, J. H. Kim, T. Hyeon, D.-H. Kim, *Nat. Commun.* **2015**, *6*, 7149.

[112] M. Singh, H. M. Haverinen, P. Dhagat, G. E. Jabbour, *Adv. Mater.* **2010**, *22*, 673-685.

[113] P. Calvert, *Chem. Mater.* **2001**, *13*, 3299-3305.

[114] J. Han, D. Ko, M. Park, J. Roh, H. Jung, Y. Lee, Y. Kwon, J. Sohn, W. K. Bae, B. D. Chin, C. Lee, *J. Soc. Inf. Disp.* **2016**, *24*, 545-551.

[115] C. Jiang, L. Mu, J. Zou, Z. He, Z. Zhong, L. Wang, M. Xu, J. Wang, J. Peng, Y. Cao, *Sci. China Chem.* **2017**, *60*, 1-7.

[116] B. H. Kim et al., *Nano Lett.* **2015**, *15*, 969-973.

[117] J. S. Park et al., *Nano Lett.* **2016**, *16*, 6946-6953.

[118] M. Layani, M. Gruchko, O. Milo, I. Balberg, D. Azulay, S. Magdassi, *ACS Nano* **2009**, *3*, 3537-3542.

[119] C. Jiang, Z. Zhong, B. Liu, Z. He, J. Zou, L. Wang, J. Wang, J. Peng, Y. Cao, *ACS Appl. Mater. Interfaces* **2016**, *8*, 26162-26168.

[120] A. Singh, R. D. Gunning, S. Ahmed, C. A. Barrett, N. J. English, J.-A. Garate, K. M. Ryan, *J. Mater. Chem.* **2012**, *22*, 1562-1569.

[121] K. M. Ryan, A. Mastroianni, K. A. Stancil, H. Liu, A. P. Alivisatos, *Nano Lett.* **2006**, *6*, 1479-1482.

[122] M. F. Toney, T. P. Russell, P. A. Logan, H. Kikuchi, J. M. Sands, S. K. Kumar, *Nature* **1995**, *374*, 709-711.

[123] Y. Amit, A. Faust, I. Lieberman, L. Yedidya, U. Banin, *Phys. Status Solidi A* **2012**, *209*, 235-242.

[124] E. Ploshnik, A. Salant, U. Banin, R. Shenhar, *Adv. Mater.* **2010**, *22*, 2774-2779.

[125] A. Persano, M. De Giorgi, A. Fiore, R. Cingolani, L. Manna, A. Cola, R. Krahne, *ACS Nano* **2010**, *4*, 1646-1652.

[126] Q. Zhang, S. Gupta, T. Emrick, T. P. Russell, *J. Am. Chem. Soc.* **2006**, *128*, 3898-3899.

[127] E. Ploshnik, A. Salant, U. Banin, R. Shenhar, *Phys. Chem. Chem. Phys.* **2010**, *12*, 11885-11893.

[128] D. Steiner, D. Azulay, A. Aharoni, A. Salant, U. Banin, O. Millo, *Phys. Rev. B* **2009**, *80*, 195308.

[129] T. Wang, X. Wang, D. Lamontagne, Z. Wang, Y. C. Cao, *J. Am. Chem. Soc.* **2013**, *135*, 6022-6025.

[130] International Telecommunication Union -ITU-R, *Recomm. ITU-R BT. 2020* **2015**, *10*, 1-8.

[131] T. Erdem, H. V. Demir, *Nanophotonics* **2013**, *2*, 57-81.

[132] R. Zhu, Z. Luo, H. Chen, Y. Dong, S.-T. Wu, *Opt. Express* **2015**, *23*, 23680.

[133] N. N. Hewa-Kasakarage, P. Z. El-Khoury, A. N. Tarnovsky, M. Kirsanova, I. Nemitz, A. Nemchinov, M. Zamkov, *ACS Nano* **2010**, *4*, 1837-1844.

[134] F. Pisanello, L. Martiradonna, G. Leménager, P. Spinicelli, A. Fiore, L. Manna, J. P. Hermier, R. Cingolani, E. Giacobino, M. De Vittorio, A. Bramati, *Appl. Phys. Lett.* **2010**, *96*, 033101.

[135] S. Coe-Sullivan, W. Liu, P. Allen, J. S. Steckel, *ECS J. Solid State Sci. Technol.* **2012**, *2*, R3026-R3030.

[136] K. Chen, H.-C. Chen, M.-H. Shih, C.-H. Wang, M. Kuo, Y.-C. Yang, C.-C. Lin, H.-C. Kuo, S. Member, *J. Lightwave Technol.* **2012**, *30*, 2256-2261.

[138] nanosys, "QDEF-Quantum Dot Pioneers," **2017**.

[138] S. Coe, W.-K. Woo, M. Bawendi, V. Bulović, *Nature* **2002**, *420*, 800-803.

[139] P. O. Anikeeva, J. E. Halpert, M. G. Bawendi, V. Bulović, *Nano Lett.* **2009**, *9*, 2532-2536.

[140] A. H. Mueller, M. A. Petruska, M. Achermann, D. J. Werder, E. A. Akhadov, D. D. Koleske, M. A. Hoffbauer, V. I. Klimov, *Nano Lett.* **2005**, *5*, 1039-1044.

[141] V. Wood, M. J. Panzer, J. E. Halpert, J.-M. Caruge, M. G. Bawendi, V. Bulović, *ACS Nano* **2009**, *3*, 3581-3586.

[142] J. M. Caruge, J. E. Halpert, V. Wood, V. Bulović, M. G. Bawendi, *Nat. Photonics* **2008**, *2*, 247-250.

[143] J. W. Stouwdam, R. A. J. Janssen, *J. Mater. Chem.* **2008**, *18*, 1889.

[144] L. Qian, Y. Zheng, J. Xue, P. H. Holloway, *Nat. Photonics* **2011**, *5*, 543-548.

[145] J. Kwak, W. K. Bae, D. Lee, I. Park, J. Lim, M. Park, H. Cho, H. Woo, D. Y. Yoon, K. Char, S. Lee, C. Lee, *Nano Lett.* **2012**, *12*, 2362-2366.

[146] Y. Yang, Y. Zheng, W. Cao, A. Titov, J. Hyvonen, J. R. Manders, J. Xue, P. H. Holloway, L. Qian, *Nat. Photonics* **2015**, *9*, 1-9.

[147] D. V. Talapin, C. B. Murray, *Science* **2005**, *310*, 86-89.

[148] M. V. Kovalenko, M. I. Bodnarchuk, J. Zaumseil, J.-S. Lee, D. V. Talapin, *J. Am. Chem. Soc.* **2010**, *132*, 10085-10092.

[149] L. Sun, J. J. Choi, D. Stachnik, A. C. Bartnik, B.-R. Hyun, G. G. Malliaras, T. Hanrath, F. W. Wise, *Nat. Nanotechnol.* **2012**, *7*, 369-373.

[150] H. Shen, W. Cao, N. T. Shewmon, C. Yang, L. S. Li, J. Xue, *Nano Lett.* **2015**, *15*, 1211-1216.

[151] D. Bozyigit, V. Wood, *MRS Bull.* **2013**, *38*, 731-736.

[152] F. García-Santamaría, Y. Chen, J. Vela, R. D. Schaller, J. A. Hollingsworth, V. I. Klimov, *Nano Lett.* **2009**, *9*, 3482-3488.

[153] F. García-Santamaría, S. Brovelli, R. Viswanatha, J. A. Hollingsworth, H. Htoon, S. A. Crooker, V. I. Klimov, *Nano Lett.* **2011**, *11*, 687-693.

[154] D. Bozyigit, O. Yarema, V. Wood, *Adv. Funct. Mater.* **2013**, *23*, 3024-3029.

[155] J. S. Steckel, P. Snee, S. Coe-Sullivan, J. P. Zimmer, J. E. Halpert, P. Anikeeva, L. A. Kim, V. Bulovic, M. G. Bawendi, *Angew. Chem. Int. Ed.* **2006**, *45*, 5796-5799; *Angew. Chem.* **2006**, *118*, 5928-5931.

[156] R. Meerheim, M. Furno, S. Hofmann, B. Lüssem, K. Leo, *Appl. Phys. Lett.* **2010**, *97*, 253305.

[157] S. Nam, N. Oh, Y. Zhai, M. Shim, *ACS Nano* **2015**, *9*, 878-885.

[158] E. Rothenberg, M. Kazes, E. Shaviv, U. Banin, *Nano Lett.* **2005**, *5*, 1581-1586.

[159] K. Park, Z. Deutsch, J. J. Li, D. Oron, S. Weiss, *ACS Nano* **2012**, *6*, 10013-10023.

[160] J. Müller, J. M. Lupton, P. G. Lagoudakis, F. Schindler, R. Koeppe, A. L. Rogach, J. Feldmann, D. V. Talapin, H. Weller, *Nano Lett.* **2005**, *5*, 2044-2049.

[161] The European parliament and the council of European union, "DIRECTIVE 2002/95/EC on the restriction of the use of certain hazardous substances in electrical and electronic equipment", **2003**, 19-23.

[162] C. Ippen, T. Greco, Y. Kim, J. Kim, M. S. Oh, C. J. Han, A. Wedel, *Org. Electron.* **2014**, *15*, 126-131.

[163] W. Ji, P. Jing, W. Xu, X. Yuan, Y. Wang, J. Zhao, A. K. Y. Jen, *Appl. Phys. Lett.* **2013**, *103*, 053106.

[164] J. Bang et al., *Chem. Mater.* **2010**, *22*, 233-240.

[165] G. Jia, U. Banin, *J. Am. Chem. Soc.* **2014**, *136*, 11121-11127.

[166] H. Li, R. Brescia, R. Krahne, G. Bertoni, M. J. P. Alcocer, C. D'Andrea, F. Scotognella, F. Tassone, M. Zanella, M. De Giorgi, L. Manna, *ACS Nano* **2012**, *6*, 1637-1647.

[167] C. Bouet, D. Laufer, B. Mahler, B. Nadal, H. Heuclin, S. Pedetti, G. Patriarche, B. Dubertret, *Chem. Mater.* **2014**, *26*, 3002-3008.

[168] A. A. Guzelian, J. E. B. Katari, A. V. Kadavanich, U. Banin, K. Hamad, E. Juban, A. P. Alivisatos, R. H. Wolters, C. C. Arnold, J. R. Heath, *J. Phys. Chem.* **1996**, *100*, 7212-7219.

[169] R. Xie, D. Battaglia, X. Peng, *J. Am. Chem. Soc.* **2007**, *129*, 15432-15433.

[170] X.Yang, D. Zhao, K. S. Leck, S. T. Tan, Y. X. Tang, J. L. Zhao, H. V. Demir, X. W. Sun, *Adv. Mater.* **2012**, *24*, 4180-4185.

[171] W. S. Song, H. S. Lee, J. C. Lee, D. S. Jang, Y. Choi, M. Choi, H. Yang, *J. Nanopart. Res.* **2013**, *15*, 1750.

[172] M. D. Tessier, D. Dupont, K. De Nolf, J. De Roo, Z. Hens, *Chem. Mater.* **2015**, *27*, 4893-4898.

[173] B. J. Beberwyck, A. P. Alivisatos, *J. Am. Chem. Soc.* **2012**, *134*, 19977-19980.

[174] L. De Trizio et al., *Chem. Mater.* **2015**, *27*, 1120-1128.

[175] B. J. Beberwyck, Y. Surendranath, A. P. Alivisatos, *J. Phys. Chem. C* **2013**, *117*, 19759-19770.

[176] J. Lim, M. Park, W. K. Bae, D. Lee, S. Lee, C. Lee, K. Char, *ACS Nano* **2013**, *7*, 9019-9026.

[177] Q. A. Akkerman, A. Genovese, C. George, M. Prato, I. Moreels, A. Casu, S. Marras, A. Curcio, A. Scarpellini, T. Pellegrino, L. Manna, V. Lesnyak, *ACS Nano* **2015**, *9*, 521-531.

[178] W. Bryks, M. Wette, N. Velez, S.-W. Hsu, A. R. Tao, *J. Am. Chem. Soc.* **2014**, *136*, 6175-6178.

[179] B. Chen, H. Zhong, W. Zhang, Z. Tan, Y. Li, C. Yu, T. Zhai, Y. Bando, S. Yang, B. Zou, *Adv. Funct. Mater.* **2012**, *22*, 2081-2088.

[180] J.-H. Kim, H. Yang, *Opt. Lett.* **2014**, *39*, 5002.

[181] L. Protesescu, S. Yakunin, M. I. Bodnarchuk, F. Krieg, R. Caputo, C. H. Hendon, R. X. Yang, A. Walsh, M. V. Kovalenko, *Nano Lett.* **2015**, *15*, 3692-3696.

[182] Q. A. Akkerman et al., *J. Am. Chem. Soc.* **2016**, *138*, 1010-1016.

[183] S. Sun, D. Yuan, Y. Xu, A. Wang, Z. Deng, *ACS Nano* **2016**, *10*, 3648-3657.

[184] X. Li, D. Yu, F. Cao, Y. Gu, Y. Wei, Y. Wu, J. Song, H. Zeng, *Adv. Funct. Mater.* **2016**, *26*, 5903-5912.

[185] Y. Bekenstein, B. A. Koscher, S. W. Eaton, P. Yang, A. P. Alivisatos, *J. Am. Chem. Soc.* **2015**, *137*, 16008-16011.

[186] S. Aharon, L. Etgar, *Nano Lett.* **2016**, *16*, 3230-3235.

[187] M. Meyns, M. Perálvarez, A. Heuer-Jungemann, W. Hertog, M. Ibáñez, R. Nafria, A. Genç, J. Arbiol, M. V. Kovalenko, J. Carreras, A. Cabot, A. G. Kanaras, *ACS Appl. Mater. Interfaces* **2016**, *8*, 19579-19586.

[188] X. Li, Y. Wu, S. Zhang, B. Cai, Y. Gu, J. Song, H. Zeng, *Adv. Funct. Mater.* **2016**, *26*, 2435-2445.

[189] T. Udayabhaskararao, M. Kazes, L. Houben, H. Lin, D. Oron, *Chem. Mater.* **2017**, *29*, 1302-1308.

[190] X. Qiu, B. Cao, S. Yuan, X. Chen, Z. Qiu, Y. Jiang, Q. Ye, H. Wang, H. Zeng, J. Liu, M. G. Kanatzidis, *Sol. Energy Mater. Sol. Cells* **2017**, *159*, 227-234.

[191] W. Ming, H. Shi, M. H. Du, *J. Mater. Chem. A* **2016**, *4*, 13852-13858.

[192] F. Giustino, H. J. Snaith, *ACS Energy Lett.* **2016**, *1*, 1233-1240.

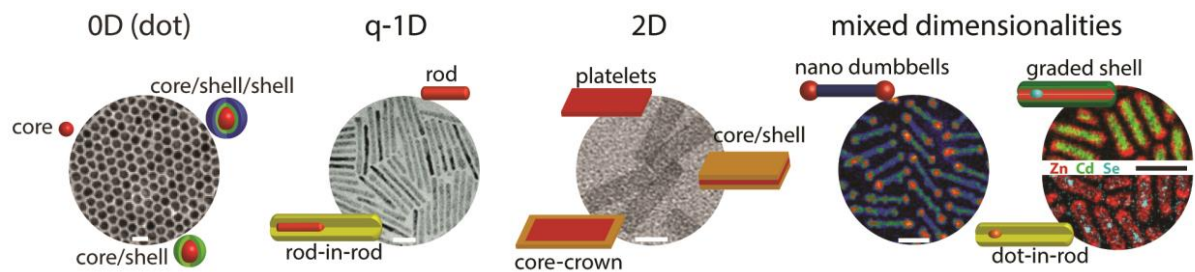

**Figure 1.** SCNCs with different dimensionalities and structures. Each dimensionality category illustrates different structures of SCNCs. The TEM image for 0D shows CdSe-core CdS/Zn$_{0.5}$Cd$_{0.5}$S/ZnS-multishell NCs (reprinted from Ref. 37 with permission). The TEM image for quasi-1D (q-1D) shows the nanorod-in-rod structure of CdSe-in-CdS NCs (reprinted from Ref. 38 with permission). TEM image for 2D presents CdSe core nanoplatelets (reprinted from Ref. 39 with permission). The left TEM image of mixed dimensionalities shows double heterojunction CdS nanorods with CdSe/ZnS core-shell tips (reprinted from Ref. 40 with permission). The right STEM image of mixed dimensionalities shows elemental mapping of CdSe/Cd$_{1-x}$Zn$_x$Se seeded nanorods with a graded shell; the upper semicircle is an overlay of Zn and Cd spatial distributions, which demonstrates the graded-shell structure, while the bottom semicircle is an overlay of Zn and Se spatial distributions, which demonstrates the seeded-rod structure (reprinted from Ref. 41 with permission). All scale bars correspond to 20 nm.

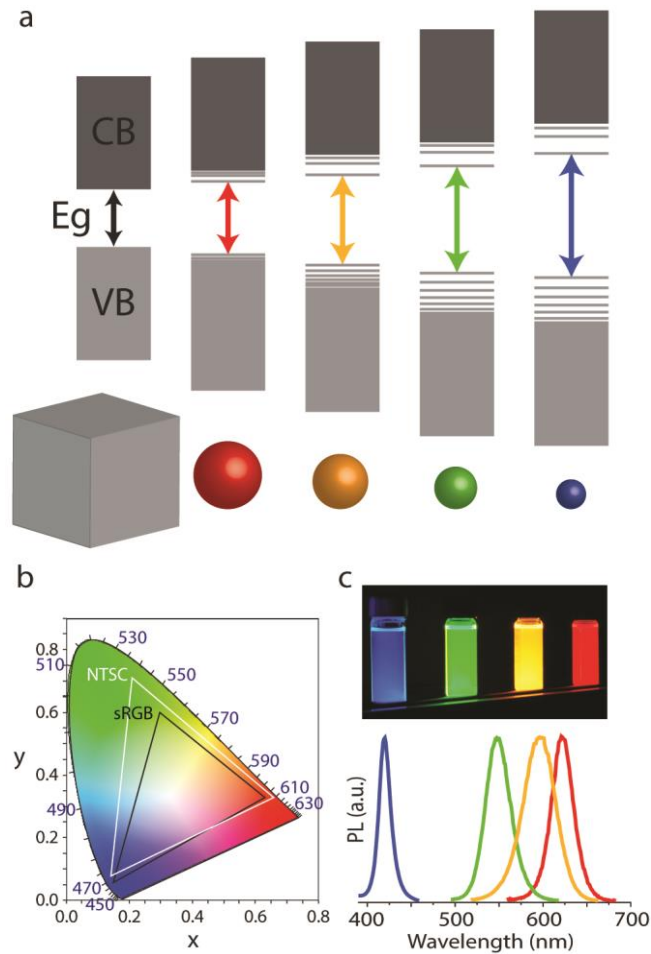

**Figure 2.** a) Schematic representation of the electronic energy structure of different-sized SCNCs, from bulk to CQDs with decreasing size. b) 1931 CIE chromaticity diagram with the s-RGB (black triangle) and the NTSC 1987 (white triangle) color gamut. The OSRAM Color Calculator software ( http://www.osram.us ) was used to map the colors. c) Induced fluorescence (UV) of SCNCs with different sizes and dimensions along with their corresponding narrow PL spectra.

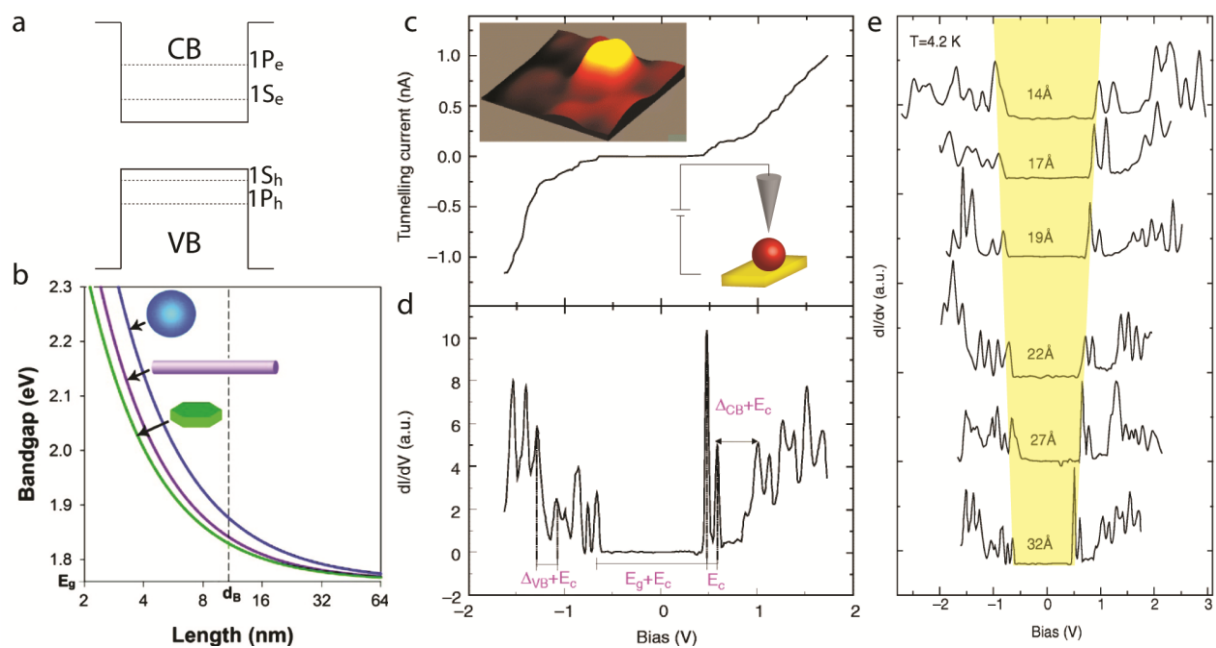

**Figure 3.** a) Schematic diagram of the conduction and valence band structure in a CQD. b) Effect of the shape on the electronic properties of SCNCs. The diagram presents the $E_g$ value of CdSe quantum wells, wires, and dots, as a function of the length of the confined dimension. The dotted line represents the exciton Bohr radius. Reprinted from Ref. 60 with permission. c) Tunneling *I-V* curve of InAs QDs exhibiting single-electron tunneling effects. The left inset presents a 10×10 nm$^2$ STM topographic image of the QD. The right inset shows a schematic illustration of the STM measurement; the QD is linked to the surface, while the tip is positioned above the QD. d) Tunneling conductance spectrum (d*I*/d*V* versus *V*). The arrows indicate the main energy separations: $E_c$ is the single-electron charging energy, $E_g$ is the nanocrystal band gap, and $\Delta VB$ and $\Delta CB$ are the spacings between levels in the valence and conduction bands, respectively. e) STS spectra of a size series of InAs CQDs. As the size of the CQD is decreased, the energy gap increases. (c)-(e) reprinted from Ref. 62 with permission.

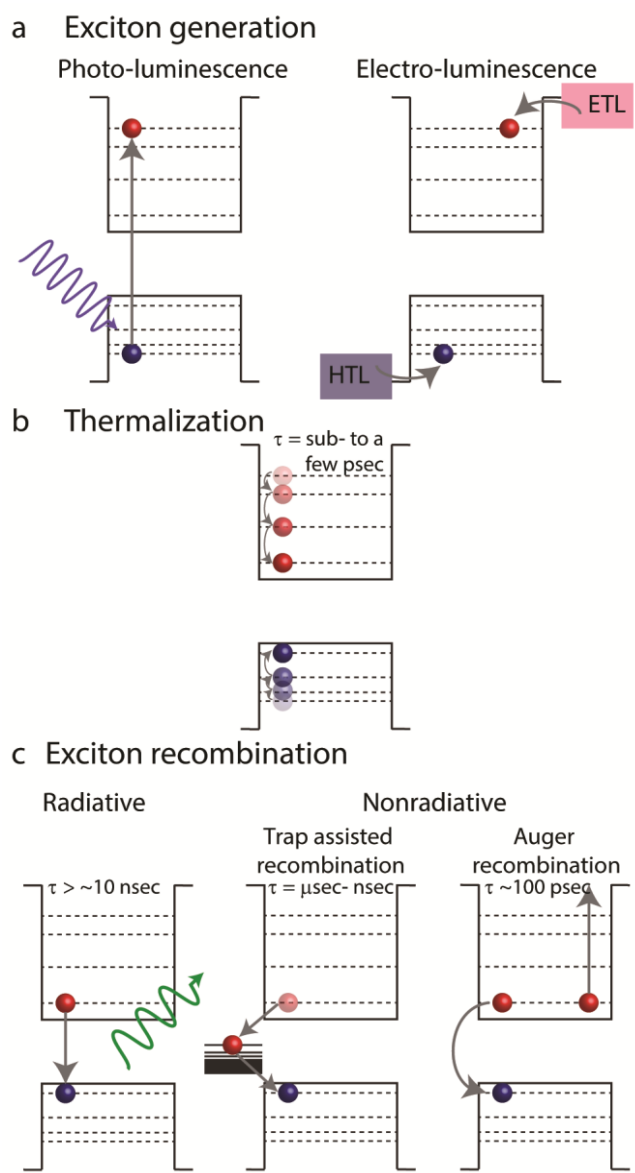

**Figure 4.** Schematic diagrams representing the dynamics of the charge carriers in SCNCs. a) Two ways for generating an exciton: by photon absorption, in which an electron is excited to the conduction band, thereby leaving a hole in the valence band, or by injecting electrons and holes into the conduction band and valence band, respectively. b) A thermalization process, in which a high-energy exciton decays to the band gap. c) Exciton recombination routes: radiatively, by emitting a photon with an energy equivalent to the band gap; nonradiatively, by decaying to a trap state; or transferring the band gap energy to excite a third charge carrier (Auger recombination).

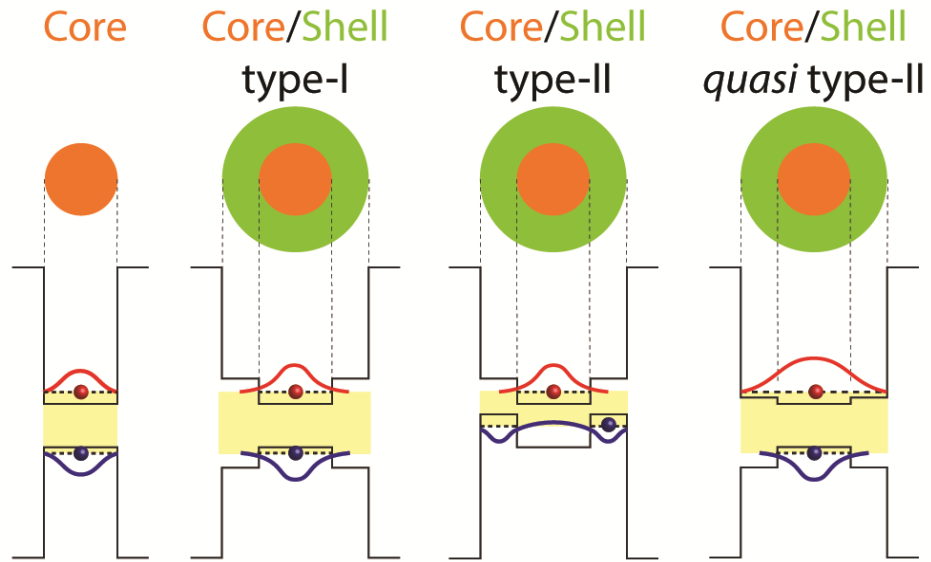

**Figure 5.** Schematic description of different band alignment possibilities in core-shell heterostructures. The resultant electron (red) and hole (blue) wave functions are illustrated.

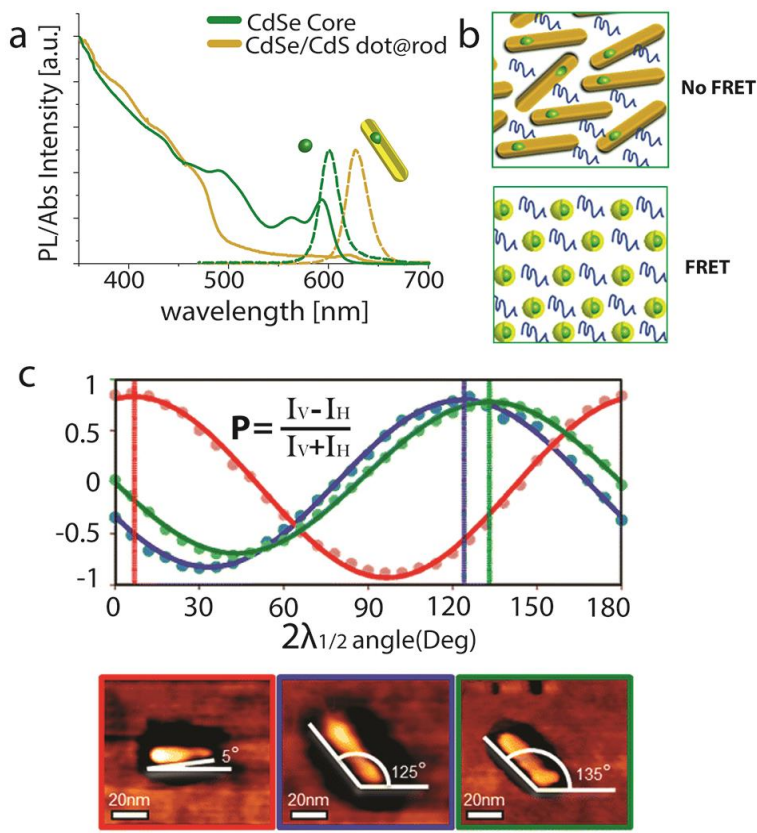

**Figure 6.** a) Absorption and PL emission spectra of bare CdSe seed CQDs (4.3 nm in diamater, green) and the corresponding spectra of CdSe/CdS-seeded rods with the same seeds (dimensions 4.7×30 nm, yellow), showing the separation between the absorption and emission upon rod growth. b) Illustration of close-packed SCNCs layer. Whereas the spatial separation between the emitting cores is short for dot-in-sphere structures and leads to a FRET process, the emitting cores for dot-in-rod structures are separated and FRET is eliminated. Reprinted from Ref. 85 with permission). c) Emission polarization (P) of several dot-in-rod SCNCs correlated with their AFM image. The polarization direction is aligned with the long axis of the rod. Reprinted from ref. 86 with permission.

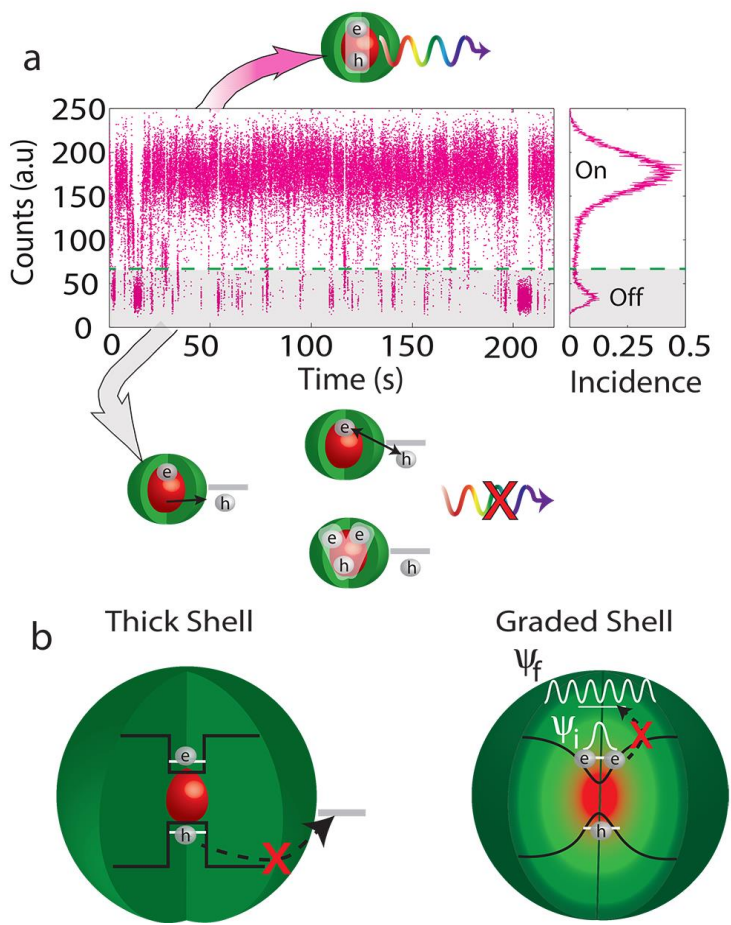

**Figure 7.** a) Typical emission time trajectory from a single NC. The intensity flickers between "on" and "off" states. Whereas every electron-hole pair recombine radiatively during the "on" state, the ejection of one of the charge carriers to a surface trap results in the SCNC moving into the "off" state, where the exciton recombines through a trap state, or alternatively the core is left charged, which leads to an enhanced nonradiative Auger process with another exciton. b) Two ways to eliminate blinking: either by growing a thick shell which prevents tunneling of one of the charge carriers to surface traps or by growing a graded shell which suppresses the Auger process.

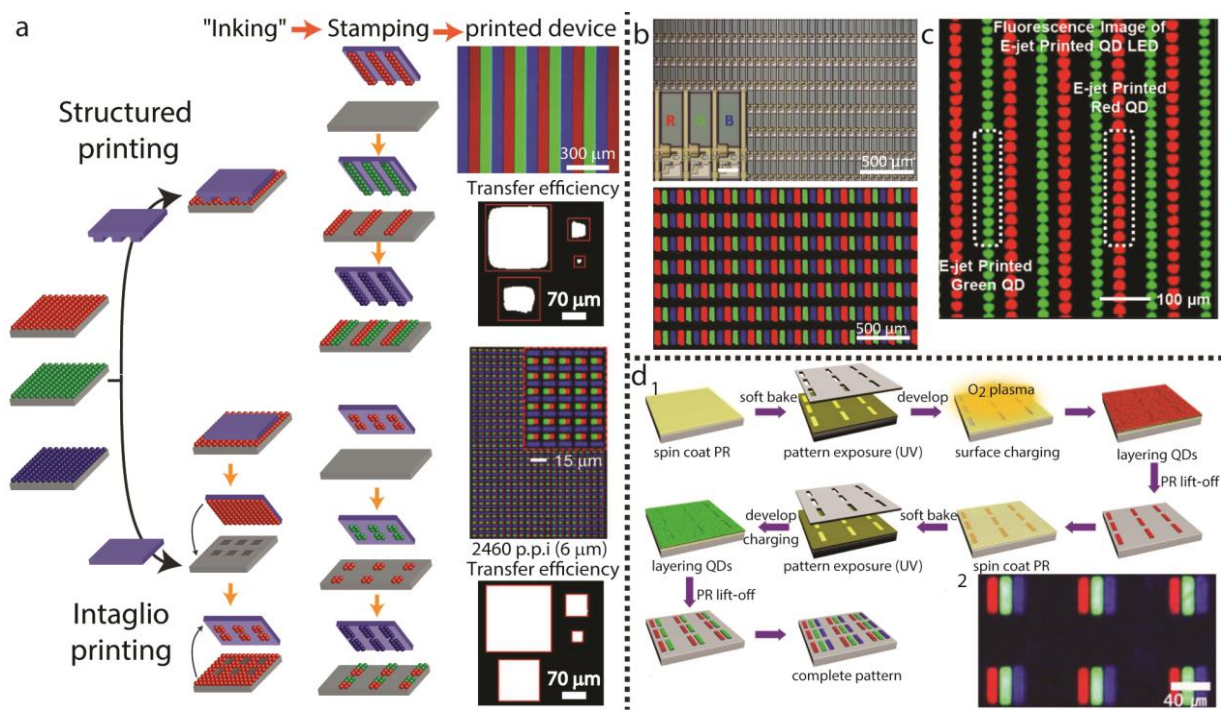

**Figure 8.** Assembly techniques for display applications. a) Two mechanisms for transfer printing are illustrated: structured printing, in which the stamp replicates its pattern to the substrate upon stamping, and intaglio printing, in which the "inked" patternless stamp is pressed onto an engraved surface, to retain its "negative" pattern, which is later transferred to the substrate. Both methods lead to full-color CQDs patterns. The upper image above the printed device column for each printing method depicts PL images of the transfer-printed RGB QD patterns (reprinted from Refs. 110 and 111 with permission), while the lower part shows the transfer efficiencies of a pattern with different sizes, comparing the resolution limitations of both methods (reprinted from Ref. 111 with permission). b) A full-color QD active matrix display fabricated by ink-jet printing. The top panel shows microscopy images of the pixel arrays. Inset: magnified image of RGB subpixels; the lower panel shows an EL image of RGB subpixel arrays. Reprinted from Ref. 115 with permission. c) A PL image of green and red QD arrays printed by e-jet printing (reprinted from Ref. 116 with permission). d) 1. Illustration of a patterning technique involving alternating photolithography and LBL assembly steps for the patterning of R, G, and B CQDs. 2. PL image of the patterned substrate (reprinted from Ref. 117 with permission).

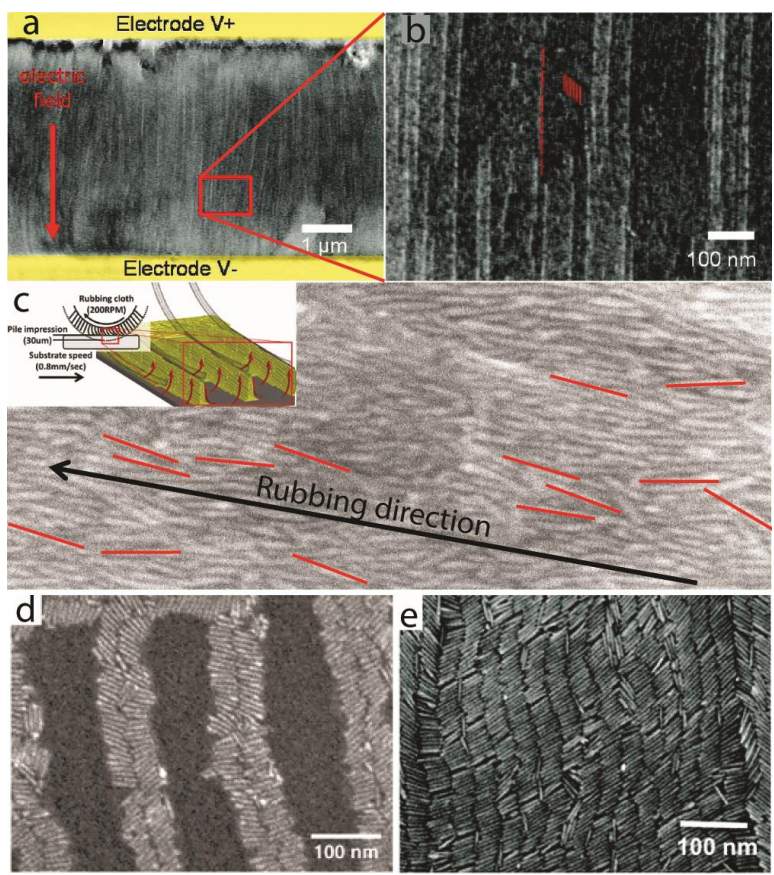

**Figure 9.** Unidirectional alignment methods. a) Alignment of CdSe/CdS nanorods by applying an electric field during evaporation of the solution. The red arrow marks the direction of the electric field. b) Higher magnification of the red square in (a), where the individual nanorods are resolved; some are highlighted in red as a guide to the eye (reprinted from Ref. 83 with permission). c) A UHR-SEM image of a thin film with aligned rods on top, obtained by rubbing with a cloth. The rods clearly demonstrate a preferred orientation along the rubbing direction. Some rods are colored in red as a guide to the eye. Inset: Illustration of the mechanical rubbing process: a rubbing cloth is pressed onto a spin-coated film; the rotation of the sample in the opposite direction to the rotation of the fibers causes the rods to pile up along the trenches of the fibers, and form ordered arrays of aligned rods (reprinted from Ref. 123 with permission). d) SEM image of aligned CdSe nanorods along a block copolymer domain (reprinted from Ref. 124 with permission). e) SEM image of CdSe/CdS nanorods aligned by applying a voltage while slow evaporation along an interface occurs. In this method, ribbons of NRs oriented parallel to the electric field are obtained (reprinted from Ref. 125 with permission).

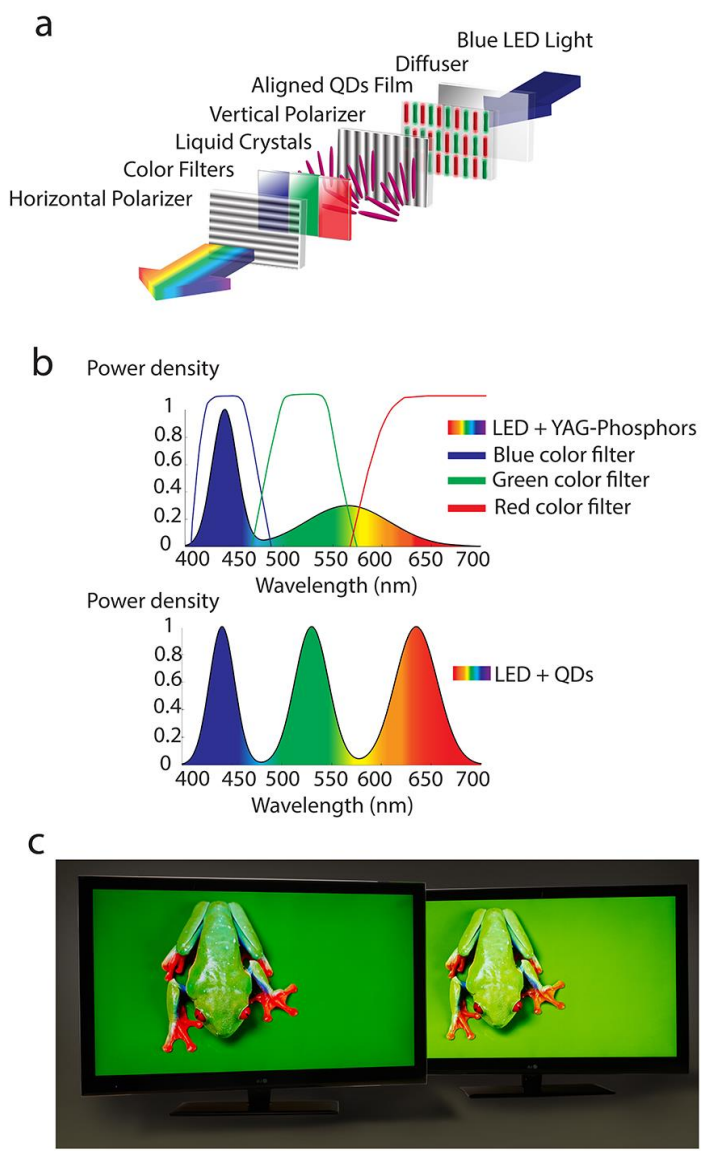

**Figure 10.** a) Typical architecture of a LCD display containing SCNCs. In this scheme, the anisotropic NCs are aligned with the vertical polarizer, thereby leading to an improvement of the efficiency. b) Schematic spectrum of a BLU of GaN LEDs with YAG phosphors along with the red, green, and blue color filters compared to a BLU containing SCNCs which can be tuned to match the color filters. c) Comparison of two LCD TV displays with and without a film containing SCNCs (left and right displays, respectively), showing the expanded color gamut which leads to a brighter, efficient, and livelier picture with SCNCs (reproduced from http://www.nanosysinc.com/media-room/media/ with permission).

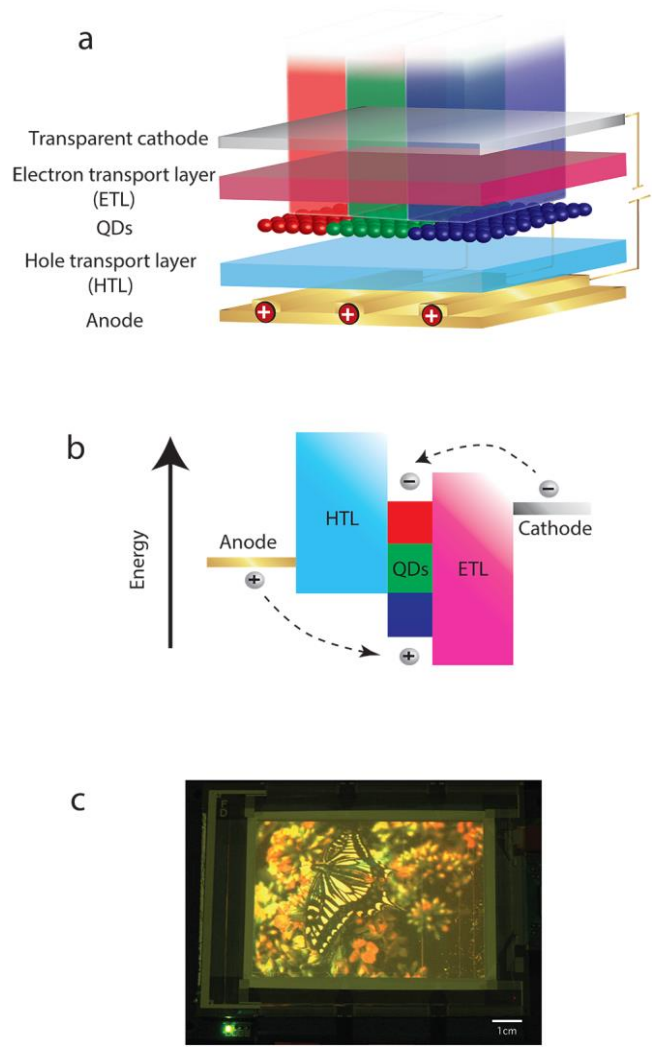

**Figure 11.** a,b) Typical architecture of EL QD-LED display layers along with their band alignment. Electrons and holes are injected via the transport layers into the emissive layer where they recombine radiatively. c) The first demonstration of a full-color 4-inch EL QD-LED display, where the pixels were patterned by a microcontact printing method with a resolution of up to 1000 pixels per inch (reprinted from Ref. 110 with permission).

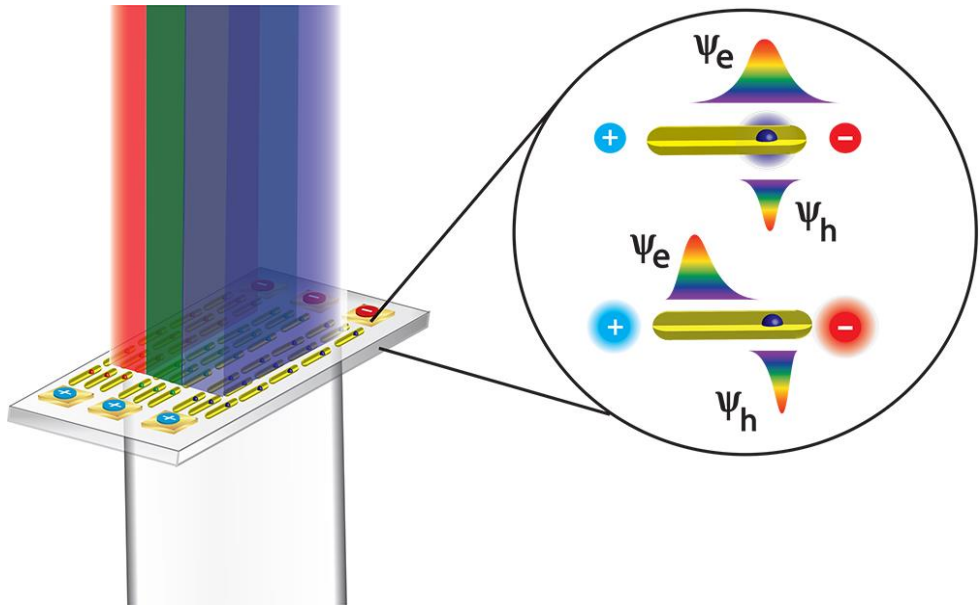

**Figure 12.** A new possible mode of operation for SCNCs in a display where the dot-in-rod is aligned with the electric field lines in the red, green, and blue subpixels. Application of an electric field bias causes the electron and hole wave functions to be spatially separated, thus enabling intensity switching to produce any color from a combination of red, green, and blue base colors.

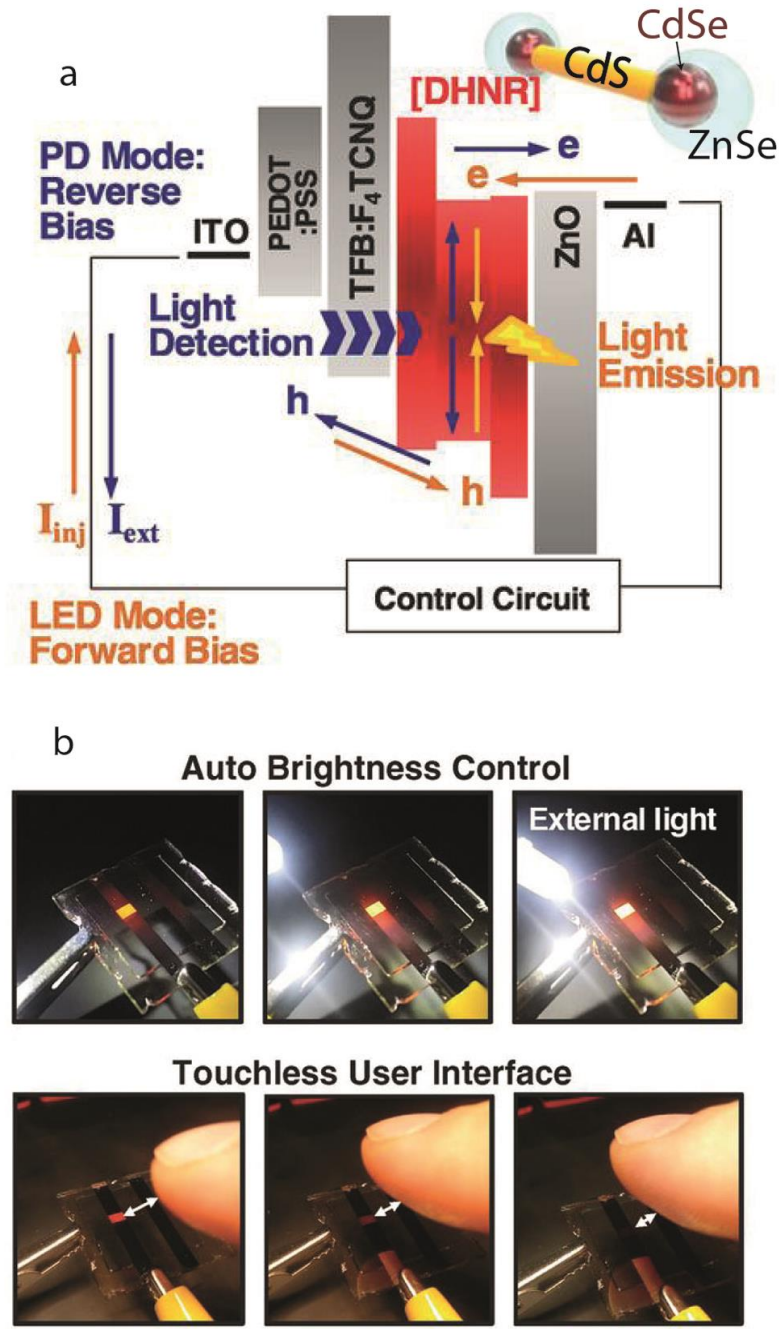

**Figure 13.** New functionalities for SCNC displays. a) Energy band diagram of a DHNR-LED along with directions of charge flow for the emission (orange arrows) and detection (blue arrows) of light as well as a schematic representation of a DHNR. b) Automatic brightness control at the single-pixel level in response to an approaching white LED bulb or an approaching finger that blocks ambient light, thus paving the way towards interactive displays.[12]

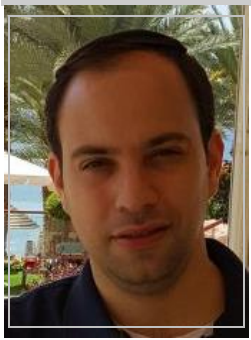

*Yossef E. Panfil completed his BSc at JCT and then his MSc in Applied Physics at the Hebrew University of Jerusalem. He is currently a PhD student at the Hebrew University of Jerusalem in Prof. Banin's group. His research focuses on the study of excitons and multiexcitons behavior, interaction, and control in a variety of colloidal semiconductor nanostructures by using single nanocrystal spectroscopy.*

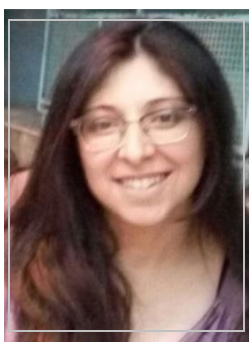

*Meirav Oded received her PhD (2016) in chemistry from the Hebrew University of Jerusalem under the guidance of Prof. Roy Shenhar. During her PhD she studied the selective deposition of polyelectrolytes on block copolymer templates with nanometric resolution. She is currently a staff scientist in the group of Prof. Banin. Her research interests are surface chemistry and self-assembly methods of colloidal semiconductor nanocrystals.*

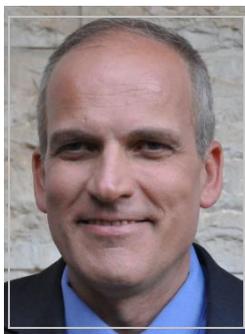

*Uri Banin holds the Alfred & Erica Larisch Memorial Chair at the Institute of Chemistry and the Center for Nanoscience and Nanotechnology at the Hebrew University of Jerusalem (HU). He was the founding director of the HU Center for Nanoscience and Nanotechnology (2001-2010) and is a founder of Qlight Nanotech, which was acquired by Merck KgaA. His research focuses on chemistry and physics of nanocrystals including synthesis of nanoparticles, size- and shape-dependent properties, and emerging applications in displays, lighting, solar energy harvesting, 3D printing, electronics, and biology.*